\documentclass[usenatbib]{article}
\usepackage{times,graphicx,a4wide,amsmath,natbib}
\pdfoutput=1
\newcommand{\msol}{M_{\rm \odot}}
\newcommand{\mjup}{M_{\rm Jup}}

\begin{document}
	
\title{Surface Flux Patterns on Planets in Circumbinary Systems, and Potential for Photosynthesis}
\author{Duncan H. Forgan$^{1,2}$, Alexander Mead $^1$, Charles S. Cockell $^2$, John A. Raven $^3$}
\maketitle

\noindent $^{1}$Scottish Universities Physics Alliance (SUPA), Institute for Astronomy, University of Edinburgh, Blackford Hill, Edinburgh, EH9 3HJ, UK \\
$^{2}$ UK Centre for Astrobiology, School of Physics and Astronomy, University of Edinburgh \\
$^{3}$ Division of Plant Sciences, University of Dundee at TJHI, The James Hutton Institute, Invergowrie, Dundee, UK \\

\noindent \textbf{First Draft} \\
\noindent \textbf{Word Count: 5700} \\

\noindent \textbf{Direct Correspondence to:} \\
D.H. Forgan \\ \\
\textbf{Email:} dhf3@st-andrews.ac.uk \\

\newpage

\begin{abstract}

Recently, the Kepler Space Telescope has detected several planets in orbit around a close binary star system.  These so-called circumbinary planets will experience non-trivial spatial and temporal distributions of radiative flux on their surfaces, with features not seen in their single-star orbiting counterparts.  Earthlike circumbinary planets inhabited by photosynthetic organisms will be forced to adapt to these unusual flux patterns.

We map the flux received by putative Earthlike planets (as a function of surface latitude/longitude and time) orbiting the binary star systems Kepler-16 and Kepler-47, two star systems which already boast circumbinary exoplanet detections.  The longitudinal and latitudinal distribution of flux is sensitive to the centre of mass motion of the binary, and the relative orbital phases of the binary and planet. Total eclipses of the secondary by the primary, as well as partial eclipses of the primary by the secondary add an extra forcing term to the system.  We also find that the patterns of darkness on the surface are equally unique.  Beyond the planet's polar circles, the surface spends a significantly longer time in darkness than latitudes around the equator, due to the stars' motions delaying the first sunrise of spring (or hastening the last sunset of autumn).  In the case of Kepler-47, we also find a weak longitudinal dependence for darkness, but this effect tends to average out if considered over many orbits.

In the light of these flux and darkness patterns, we consider and discuss the prospects and challenges for photosynthetic organisms, using terrestrial analogues as a guide.

\end{abstract}


\section{Introduction}\label{sec:introduction}

Extrasolar planets (or exoplanets) show a rich variety of orbital architectures, challenging planet formation theory.  Their properties force astrophysicists and astrobiologists to re-examine their assumptions regarding the growth and evolution of potentially habitable planets in the Milky Way.  

Typically, the astrophysical ``habitability'' of a world is determined by the radiative flux received from an external radiation source.  This flux is used as input for climate models, which simulate the response of a planet of Earth mass - with similar composition and atmosphere - to this radiation.  This allows the construction of a habitable zone (HZ), a region surrounding the radiation source that should be capable of hosting planets of Earth mass and composition with liquid water on their surface.

This habitable zone concept pre-dates the detection of exoplanets by several decades \citep{Huang1959, Hart_HZ}, culminating in the seminal 1D radiative transfer calculations of \citet{Kasting_et_al_93}, which defined inner and outer radial boundaries of a spherically symmetric HZ, based on the star's luminosity and effective temperature.  These calculations would form the basis for a number of parametrisations of the HZ \citep{Underwood2003, Selsis2007, Kaltenegger2011}.  Recently, \citet{Kopparapu2013} returned to Kasting et al's original calculations, updating the atmospheric absorption models to refine the inner and outer boundaries of the HZ, extending them (in the conservative case) to 0.99 and 1.7 AU respectively, although these boundaries apply only to Earth mass planets - increasing the mass tends to move the inner boundary closer to the star \citep{Kopparapu2014}.

The HZ model has been strongly challenged since the first detection of an exoplanet in a main sequence star system  \citep{Mayor1995}.  For example, the orbital eccentricity of exoplanets is often much higher than exhibited in the Solar System, allowing planets to move in and out of the HZ over the course of an orbit.  The habitability of a world then becomes a function of the length of time it spends inside the zone \citep{Williams2002,Kane2012,Kane2012a}.  

The advent of the Kepler Space Telescope has allowed, for the first time, meaningful statistical discussion of how frequently stars in the Milky Way possess planets in the HZ \citep{Dressing2013,Petigura2013,Kasting2013}.  These analyses are generally sensitive to properties such as stellar mass, and the precise definition of the inner and outer HZ boundaries, which are in turn sensitive to properties such as the planet's surface liquid water fraction and atmospheric composition.  Nevertheless, the above authors find the frequency of ``Earth-like'' planets around other stars to range from about 15\% to as high as 50\%.

One of Kepler's achievements which deserves note is the first detection of planetary systems orbiting binary main sequence stars.  These possess planets in orbit of the centre of mass of a binary system.  They are commonly referred to as P-type or circumbinary systems, in contrast to S-type systems where planets orbit a single component of the binary.  The first system to be detected was Kepler-16b \citep{Doyle2011}, followed quickly by Kepler-34b and Kepler-35b \citep{Welsh2012}, and two planets orbiting the star Kepler-47 \citep{Orosz2012}.  Orosz et al calculated that Kepler-47c was within the HZ of its multi-star system, although this initial calculation assumed that the radiative flux arrived from the system's centre of mass, and not the individual stars.  

As the HZ is sensitive to the flux received at the planet's surface, the HZ of circumbinary systems can differ significantly from the equivalent single-star HZ, due to a) the extra source of radiation from the companion star, and b) the extra effective sink of radiation that may come from eclipses of one star by the other.  Several groups have investigated a) in detail, revising the HZ parametrisation of \citet{Kasting_et_al_93}  for a multiple star system.  \citet{Kane2013} combined the radiation from both stars to approximate a single blackbody flux received at the planet, with the resulting effective temperature used to determine the standard HZ limits at all points in the system.  \citet{Haghighipour2013} take a similar approach, using the previous HZ limits modified to account for a second star, where the flux from each star is spectrally-weighted.  Most recently, \citet{Cuntz2014} presents a general analytic solution to solving for HZs in both S-type and P-type binary systems.  All approaches result in a HZ that deviates from the standard annular form for single star systems, with the deviation being time-dependent, and a function of the stars' orbit around the centre of mass.  \citet{Forgan2014} used one dimensional energy balance modelling to determine the habitable zone as a function of planetary orbital parameters (semi-major axis, eccentricity).  By carrying out simulations rather than relying on fitting functions, the effects of eclipses could be more naturally incorporated into the model.  These varying approaches to HZ calculation agree that, for example, Kepler-47c would be considered habitable if it were not relatively massive - consequently, it may host habitable moons (cf \citealt{Quarles2012, Heller2013a, Hinkel2013, Forgan_moon1, Kipping2013a, Forgan2014a}).

Generally speaking, these calculations have been one-dimensional, tracking the radiative response of terrestrial atmospheres as a function of atmospheric depth or latitude.  If one wishes to consider the potential nature of biomes on exoplanets, then it is important to measure both the longitudinal and latitudinal distribution of flux.  Even without the presence of a second star, the relationship between a planet's orbital period and rotation period can prove crucial in the distribution of flux, depending on the planet's obliquity and eccentricity \citep{Dobrovolskis2007,Dobrovolskis2009,Dobrovolskis2013}, and must be considered if we are to investigate for example where photosynthetic organisms might reside \citep{Brown2014}.  

Photosynthetic organisms present important biosignatures in exoplanet atmospheres \citep{Wolstencroft2002,Raven2006, Seager2013,Seager2013a} that are potentially detectable by future spectroscopy missions (although see \citealt{Livengood2011}, \citealt{Rein2014} and \citealt{Brandt2014} for the challenges involved in such an endeavour).  The distribution of photosynthetic potential on an exoplanet's surface will be an important model parameter for analysing future spectroscopic data of Earthlike planets for evidence of biological activity.

In this paper, we investigate both the latitudinal and longitudinal distribution of flux on the surface of an Earthlike planet in a circumbinary system.  We use orbital parameters from Kepler-16 and Kepler-47 as an illustration of the flux patterns already present in the HZ of known exoplanet systems.  We also investigate the time spent in darkness as a function of latitude and longitude, which will prove to be important when we consider attributes of photosynthetic organisms, such as photoperiodism.  

In section \ref{sec:method}, we outline the method by which we obtain 2D surface flux and darkness patterns; in section \ref{sec:results} we display results obtained for putative Earthlike planets orbiting Kepler-16 and Kepler-47.  In section \ref{sec:discussion} we discuss the potential for photosynthesis given the calculated flux and darkness patterns, and the dependence of these patterns on factors such as the binary's orbital parameters, and the relative phase of the planetary and binary orbits.


\section{Method}\label{sec:method}

\subsection{System setup}

The system is composed of two stars, with masses $M_1$ and $M_2$ and luminosities $L_1$ and $L_2$, orbiting each other with semi-major axis $a_{bin}$ and eccentricity $e_{bin}$, with orbital period $P_{orb,bin}$.  The planet (with mass $M_p$) performs a Keplerian orbit about the centre of mass of the system, with semi-major axis $a_p$ and eccentricity $e_p$, and orbital period $P_{orb,p}$.  The orbit may be inclined by an angle $i_p$ to the binary plane, and the obliquity of the planet's rotation relative to the binary orbital plane is $\delta_p$.  At any time during the simulation, the positions of the stars and planets are given by $\mathbf{r}_i(t)$, where $i=1,2,p$, and the relative vector 

\begin{equation} 
\mathbf{r}_{ij}  = \mathbf{r}_i-\mathbf{r}_j 
\end{equation} 

\noindent The surface of the planet is divided into cells of longitude $\Lambda$ and latitude $\Phi$.  We define the longitude of noon $\Lambda_{noon,i}$ as the longitude on the planet's surface at which star $i$ is at its maximum height, where $i=1,2$:

\begin{equation} 
\Lambda_{noon,i} = \tan^{-1}\left(\frac{\bf{r}_{pi}.\bf{\hat{y}}}{\bf{r}_{pi}.\bf{\hat{x}}} \right)
\end{equation}

%

\noindent The planet rotates with a period $P_{spin}$, and hence the apparent longitude of a fixed location on the planet surface (for a given star) varies as 

\begin{equation}
\Lambda_s(t) = \Lambda(t=0) - \Lambda_{noon,i} + \frac{2\pi t}{P_{spin}} 
\end{equation}


We fix $P_{spin}=24 h$ for all runs.  Resonances between the spin and orbital periods are clearly extremely influential on the flux distribution, and both the dynamics and habitability of planets in spin-orbit resonances around single stars have been investigated in detail \citep{Dobrovolskis2007,Dobrovolskis2009, Brown2014}.  However, it is unclear what spin-orbit resonances are available to a circumbinary planetary system, if at all (see Discussion).

\begin{figure}
\includegraphics[scale=0.4]{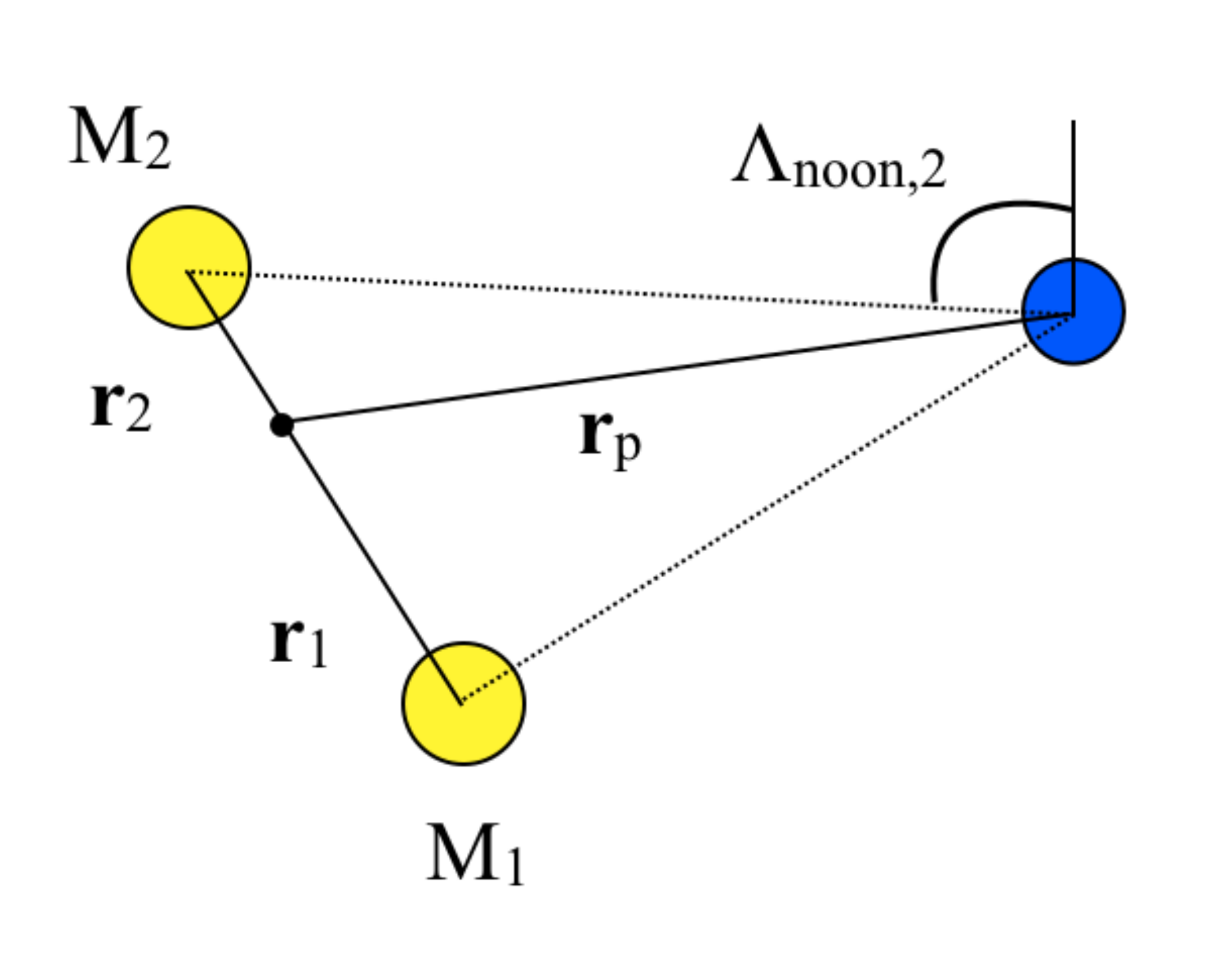} 
\caption{The simulation setup.  Stars 1 and 2 orbit the centre of mass with position vectors $\bf{r}_1$ and $\bf{r}_2$, and the planet orbits with position vector $\bf{r}_p$.  For illustration, the longitude of noon for star 2 is also shown. \label{fig:setup}}
\end{figure}

\subsection{Calculating the Received Flux}

\noindent The Cartesian surface normal vector for a given longitude/latitude cell, $\bf{n}_s(\Lambda_i(t),\Phi)$, is constructed in the customary fashion using spherical polar co-ordinates: 

\begin{align}
n_{s,x} &= \sin \Phi \cos \Lambda_i \nonumber\\
n_{s,y} &= \sin \Phi \sin \Lambda_i  \nonumber\\
n_{s,z} &= \cos \Phi \nonumber\\
\end{align}

\noindent And then rotated along the x-axis to generate the correct obliquity:

\begin{equation}
\bf{n} \rightarrow \mathbf{R}_x \bf{n}
\end{equation}

\noindent where $\mathbf{R}_x$ is the standard x-axis rotation matrix.  The flux from star $s$ hitting this latitude/longitude cell is:

\begin{equation}
F_s = \frac{L_s }{4\pi \left|\bf{r}_{ps}\right|^2}\bf{\hat{r}}_{ps}.\bf{n}_s , \,\, \mathrm{for} \,\bf{\hat{r}}_{ps}.\bf{n}_s > 0
\end{equation}

\noindent If $\bf{\hat{r}}_{ps}.\bf{n}_s \leq 0$, the latitude/longitude cell is facing away from the star, and should therefore be in its nightside.  The calculated flux is set to zero if this is the case.

\subsection{Calculating Eclipses and Periods of Darkness}

\noindent The total flux received by a latitude/longitude cell is simply

\begin{equation}
F_{tot}(\Lambda, \Phi, t) = F_1(\Lambda, \Phi, t) +F_2(\Lambda,\Phi, t)
\end{equation}

\noindent If one star eclipses the other, then the received flux must be modified.  If star $i$ eclipses star $j$, then star $j$ has its flux reduced by a factor $A_{ij}$ which is the area of intersection of two circles corresponding to the stars $i$ and $j$.

We measure the flux received at every longitude/latitude as a function of time, as well as averaging the total flux received over the course of the simulation:

\begin{equation}
\bar{F}(\Lambda,\Phi) = \frac{\int F_{tot}(\Lambda,\Phi, t') dt'}{\int dt'}
\end{equation}

\noindent This integral is evaluated as a sum with $dt\rightarrow \Delta t=30 $ mins.  We also integrate the total time spent in darkness $D_{int}$ over the course of the simulation.  We define darkness at any latitude/longitude as any time interval over which neither star is visible (i.e. $F_{tot}(t) = 0$).  We shall see that this quantity is not as straightforward as it is in single star systems.


\section{Results}\label{sec:results}

\subsection{Kepler-16b}

\noindent The Kepler-16 system consists of two relatively low-mass stars, of masses $M_1=0.6897 \msol$ and $M_2=0.2026 \msol$, orbiting each other with a semimajor axis of $a_{bin}=0.224$ AU and eccentricity $e_{bin}=0.15944$.  These parameters were taken from \citet{Doyle2011}, and we use their values for stellar radii and luminosity also.

Kepler-16b is a $0.3\mjup$ exoplanet, with a semi-major axis of 0.7 AU and eccentricity 0.069, with an orbital inclination within 0.4 deg of the binary plane.  Kepler-16 provides an example of a system with two low mass stars that will eclipse each other on a relatively long timescale (as the orbital period of the binary is 41 days), and Kepler-16b's orbit is just outside the habitable zone \citep{Kane2013,Haghighipour2013,Forgan2014}.  We will therefore replace Kepler-16b with an Earthlike planet to assess how flux is distributed in this particular scenario.

We also perform runs without the second star, where the planet orbits the primary with the same parameters.  We will use this data to identify features that are unique to circumbinary systems.   We note here that orbits at the semimajor axis of Kepler-16b are stable only for low eccentricities (see Discussion).  Some of the patterns displayed here are therefore of limited relevance to Kepler-16b in particular, but we can consider them as indicative of patterns displayed in as yet undiscovered circumbinary planetary systems where the binary has a relatively small mass ratio, while permitting stable eccentric orbits.

\subsubsection{Flux Patterns}

\noindent Figure 1 shows $\bar{F}$ on an Earthlike planet with Kepler-16b's parameters, where we run the simulation for one orbital period.  We fix the planet's semimajor axis at 0.7 AU, and vary the eccentricity between $e_p = 0,0.2,0.5$, and the obliquity between $\delta_p = 0^{\circ},30^{\circ},60^{\circ}$.  The longitudes of periapsis for both the planet and binary orbits are aligned, which we note will have an affect on resulting surface patterns. 

Generally speaking, the features produced by increasing eccentricity are similar to those in the single-star case: the flux peaks at around the equator, in this case at longitudes around 90$^\circ$.  However, in the single star case this peak appears at the substellar point of the planet when it is at periastron.  Circumbinary planets will present two substellar points at periastron - if the obliquity is small, the substellar points both appear on the equator, at slightly separated longitudes, resulting in the flux being smeared over larger angular area than in the single-star case.

If the obliquity is high, the flux is more uniformly distributed, as it is in the single star case.  However, the effect of eclipses becomes increasingly important.  The orbital period of the planet is around 200 days, around 5 full orbits of the binary.  Therefore, one half of the orbit will experience an extra eclipse period compared to the other, over the course of an orbital period.  At high obliquity, this means one polar region will receive less flux than the other - in this case, the north pole receives less flux than the south.  This is a function of the relative phases of the binary orbit and the planetary orbit - in future years this pattern will be reversed, and the north will receive more flux than the south.

\begin{figure*}
$\begin{array}{ccc}
\includegraphics[scale=0.25]{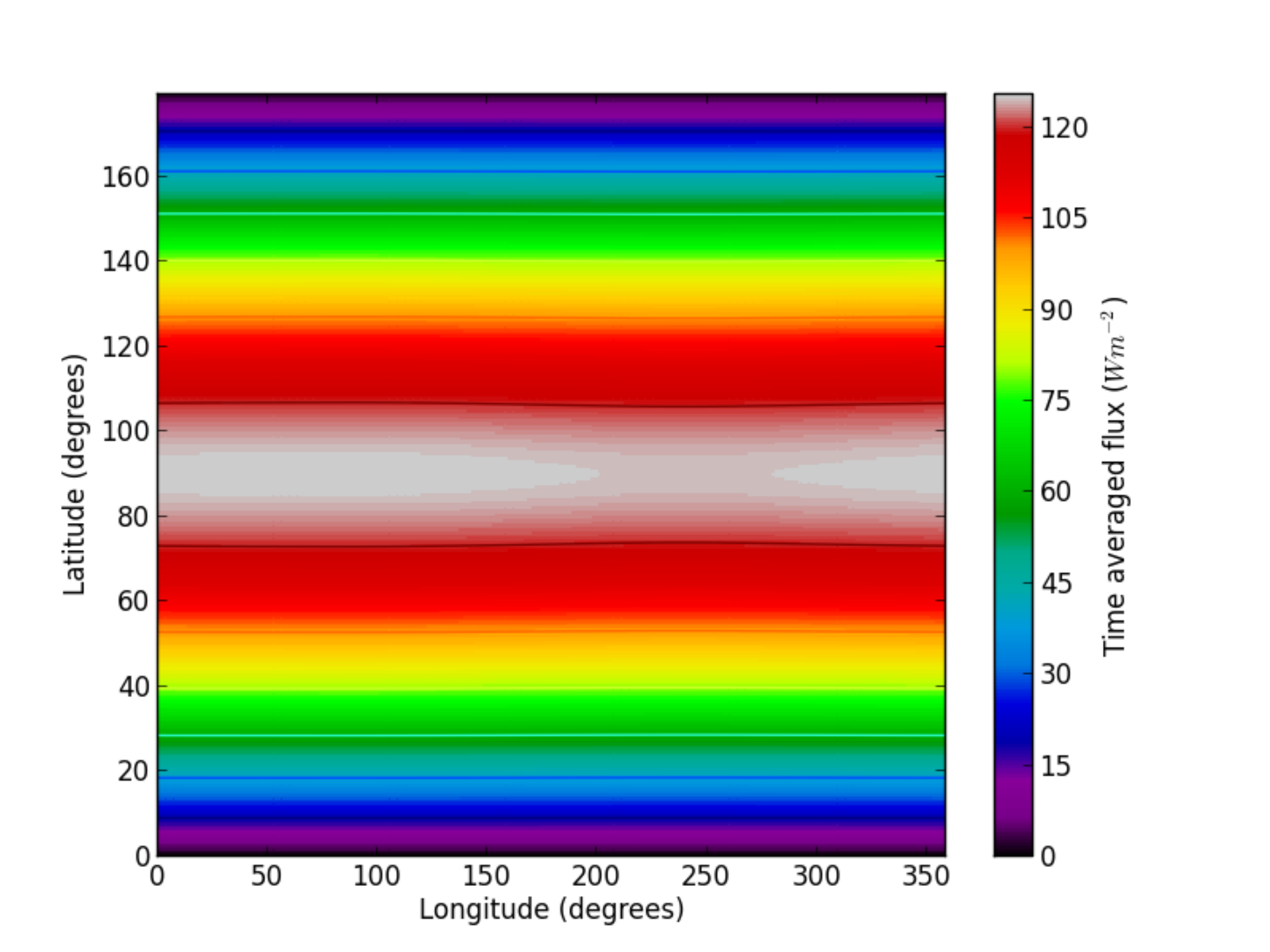} &
\includegraphics[scale=0.25]{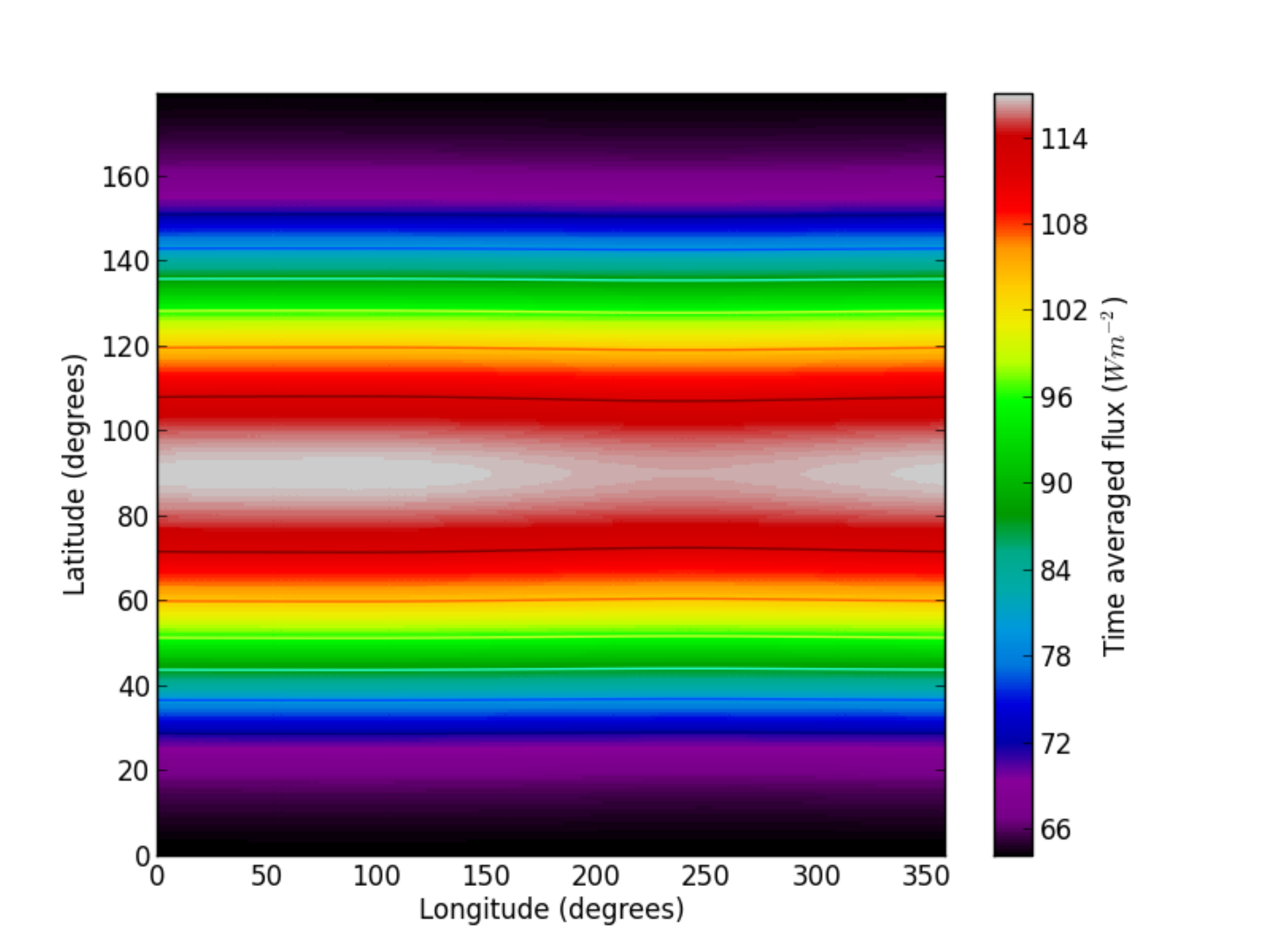} &
\includegraphics[scale=0.25]{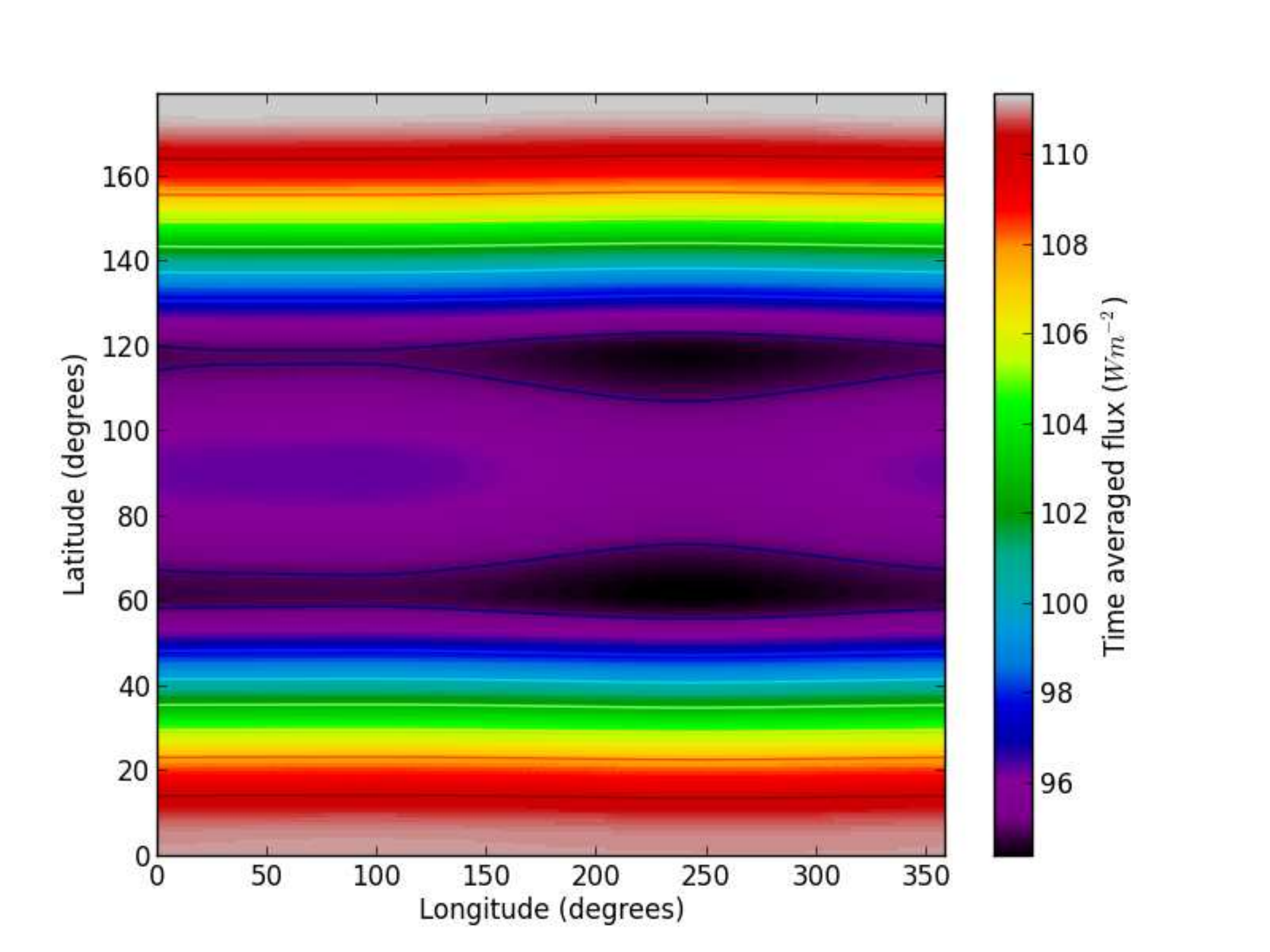} \\
\includegraphics[scale=0.25]{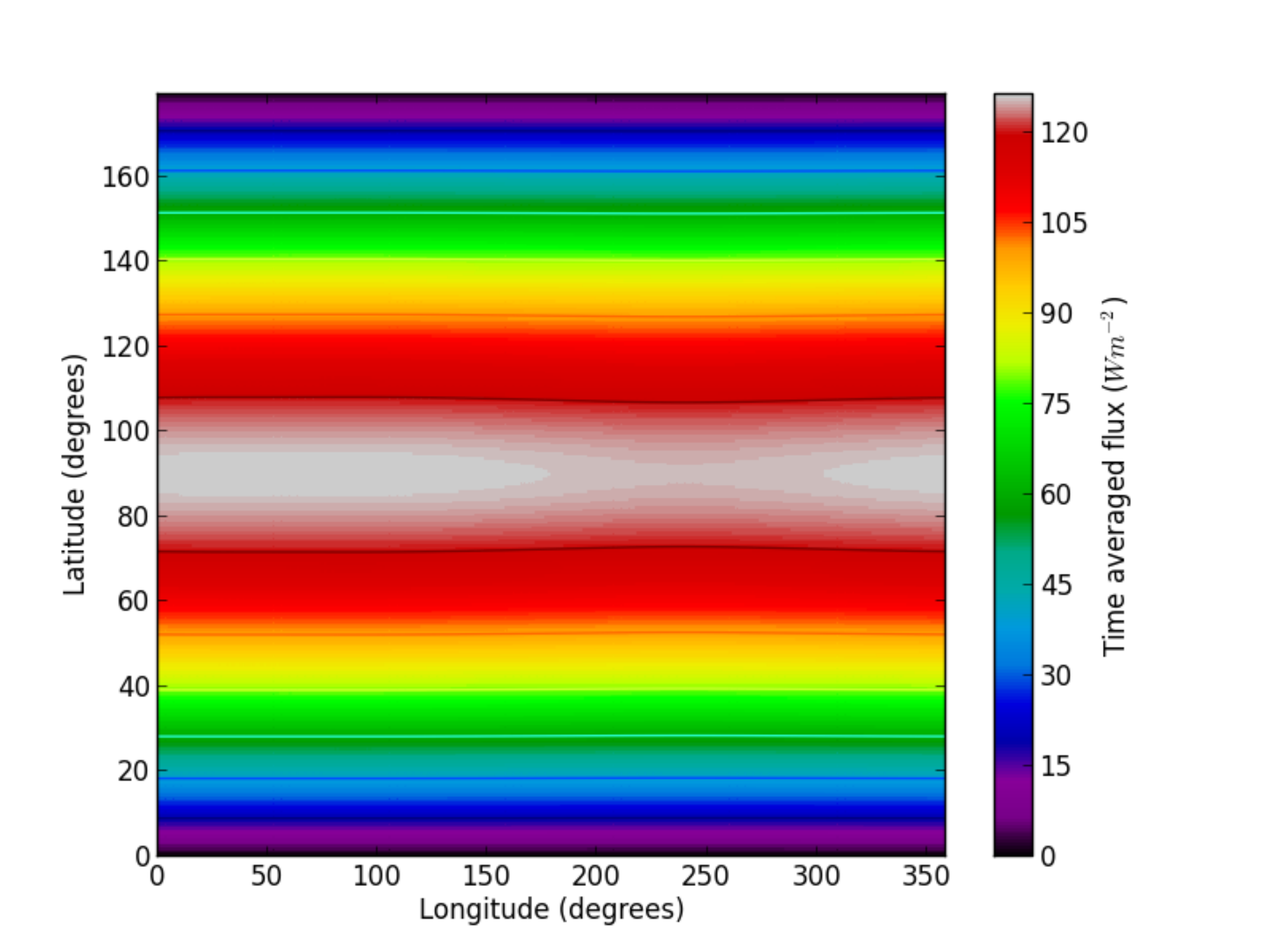} &
\includegraphics[scale=0.25]{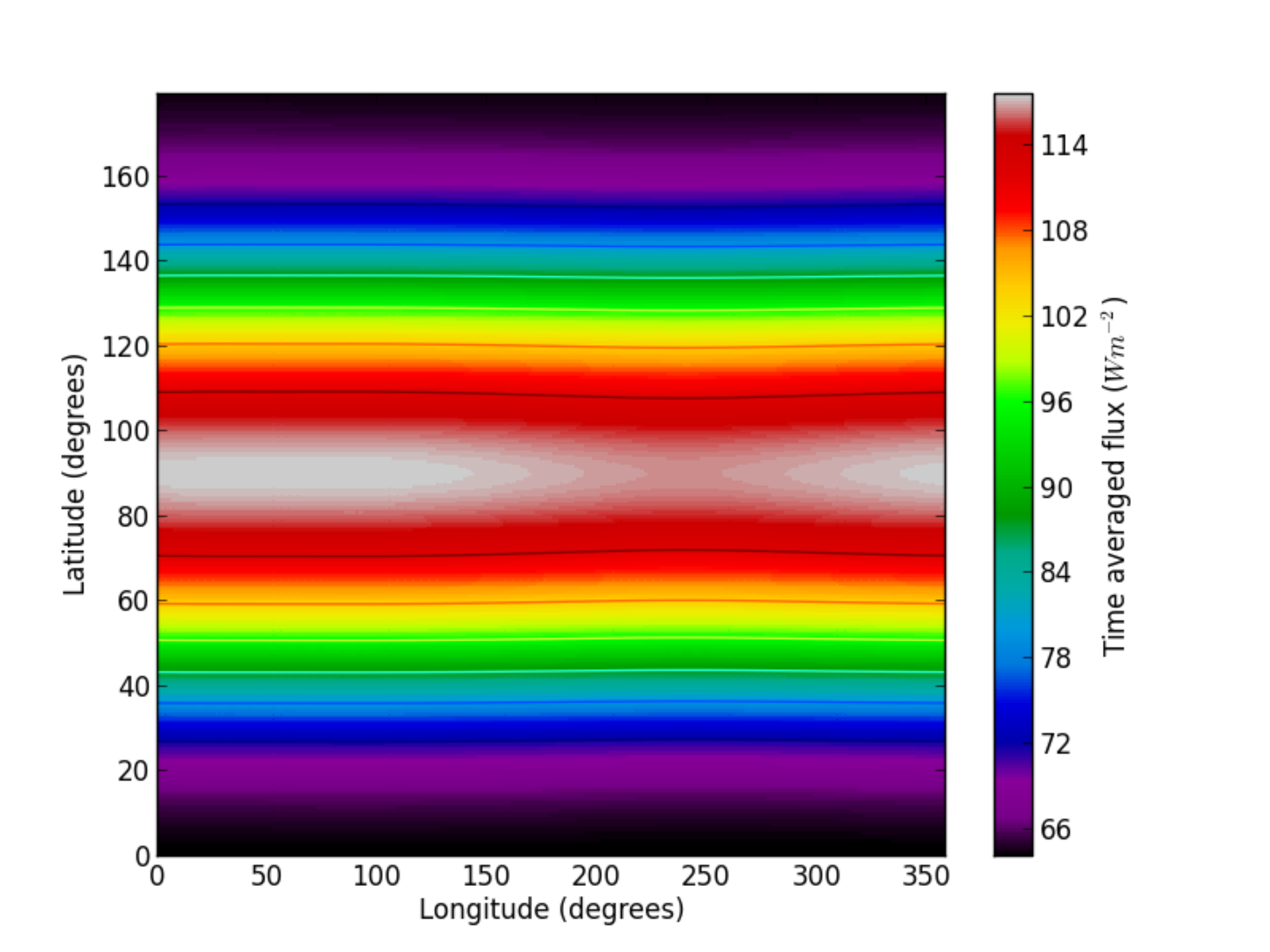} &
\includegraphics[scale=0.25]{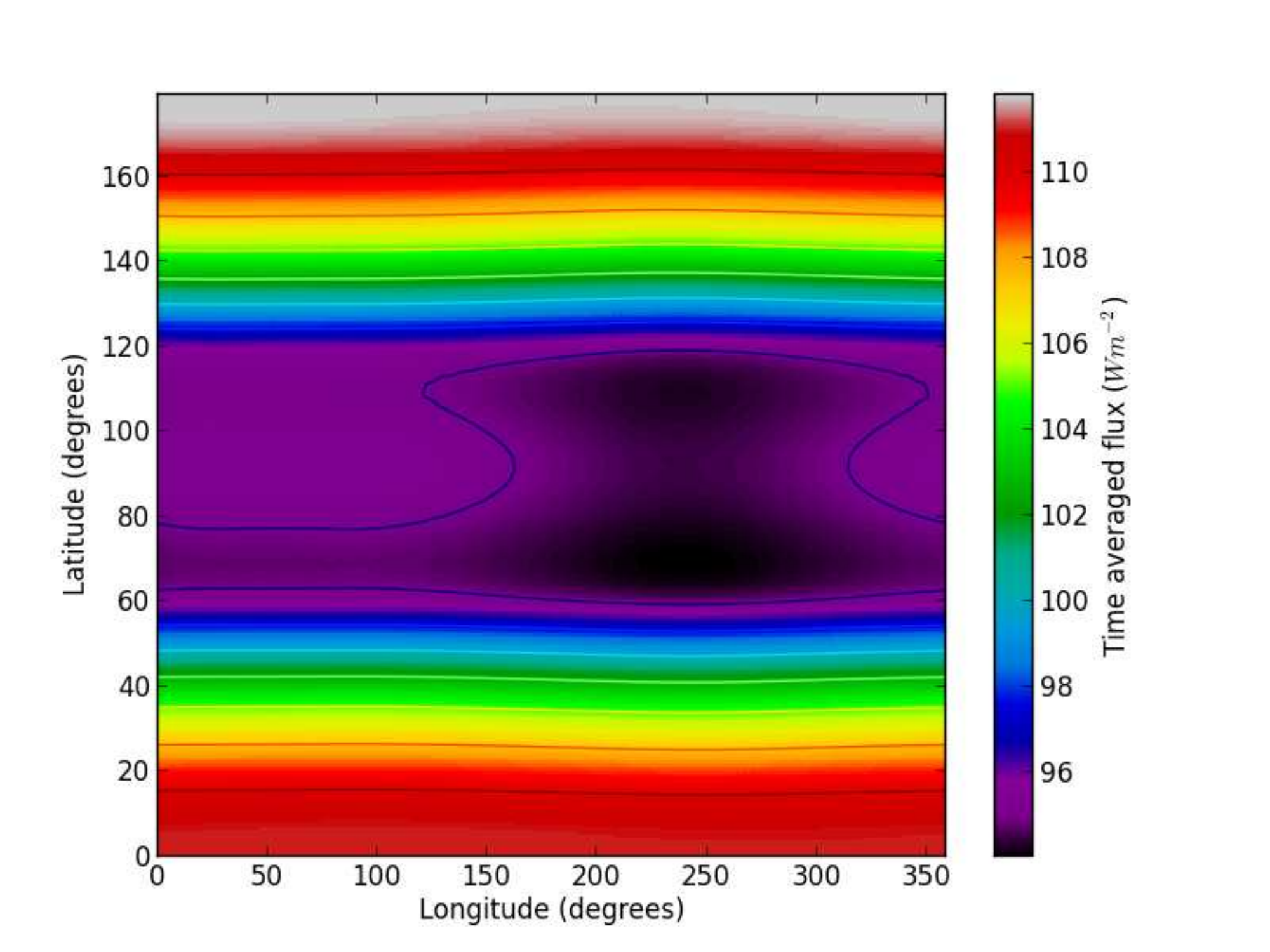} \\
\includegraphics[scale=0.25]{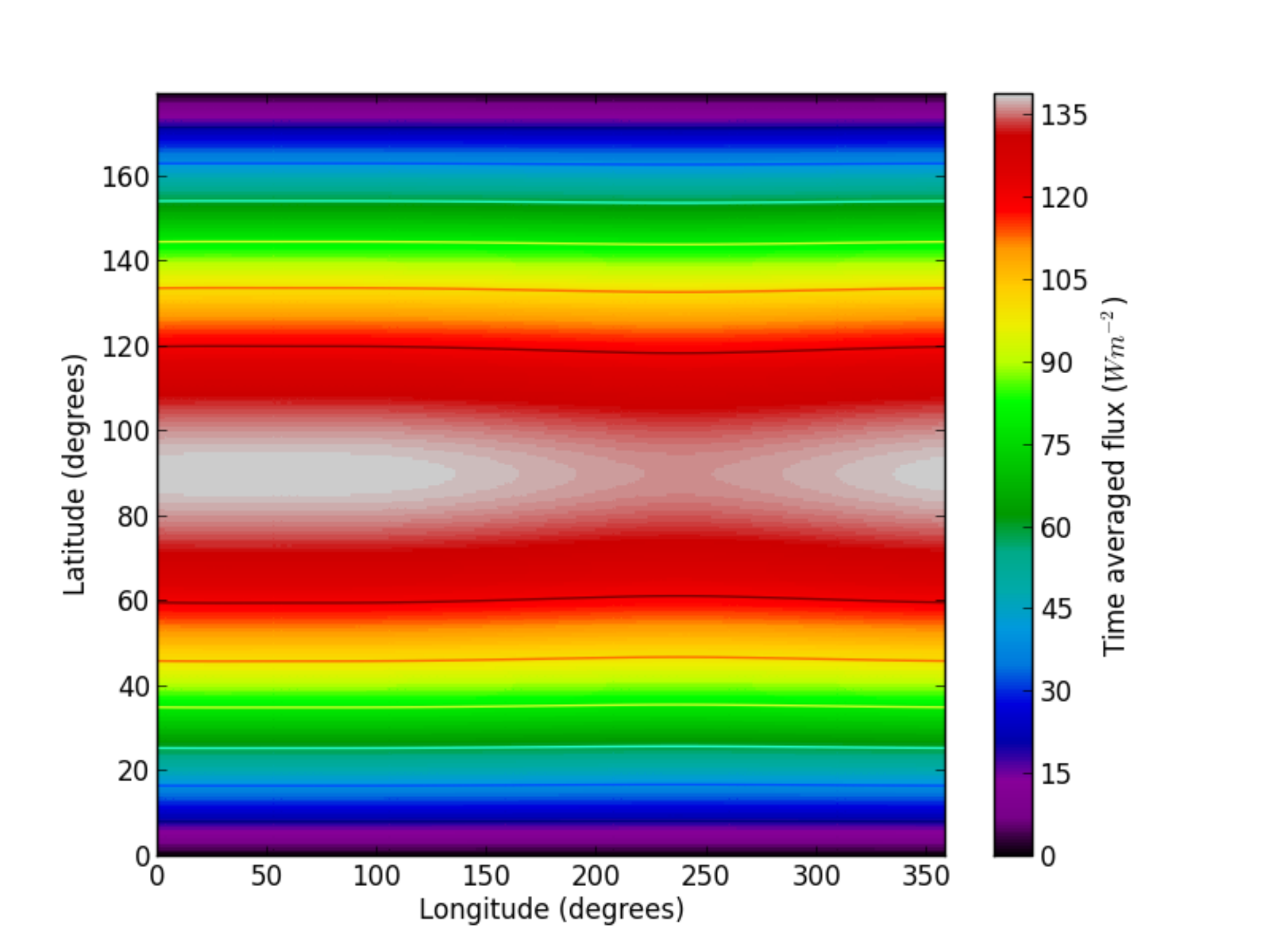} &
\includegraphics[scale=0.25]{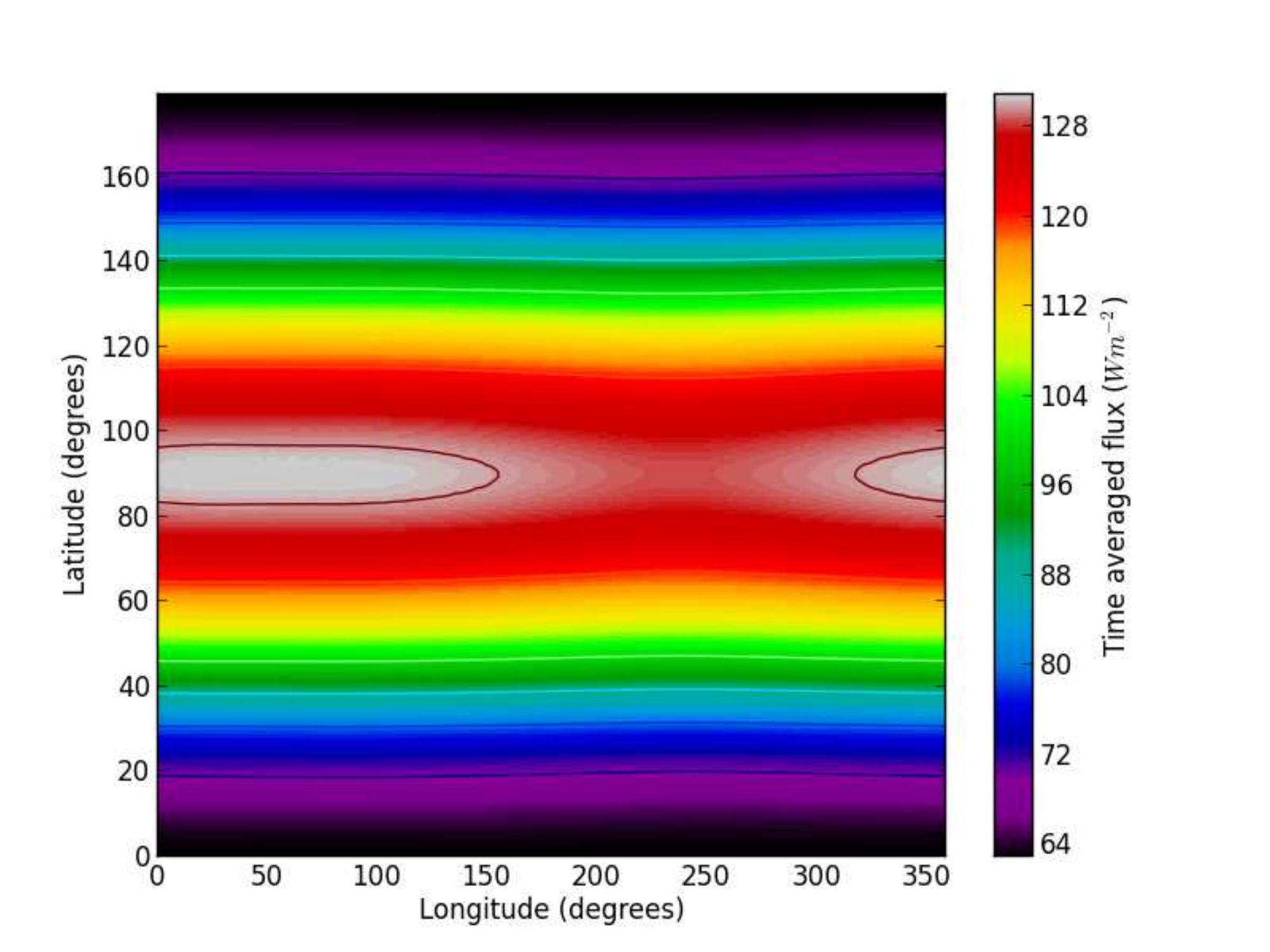} &
\includegraphics[scale=0.25]{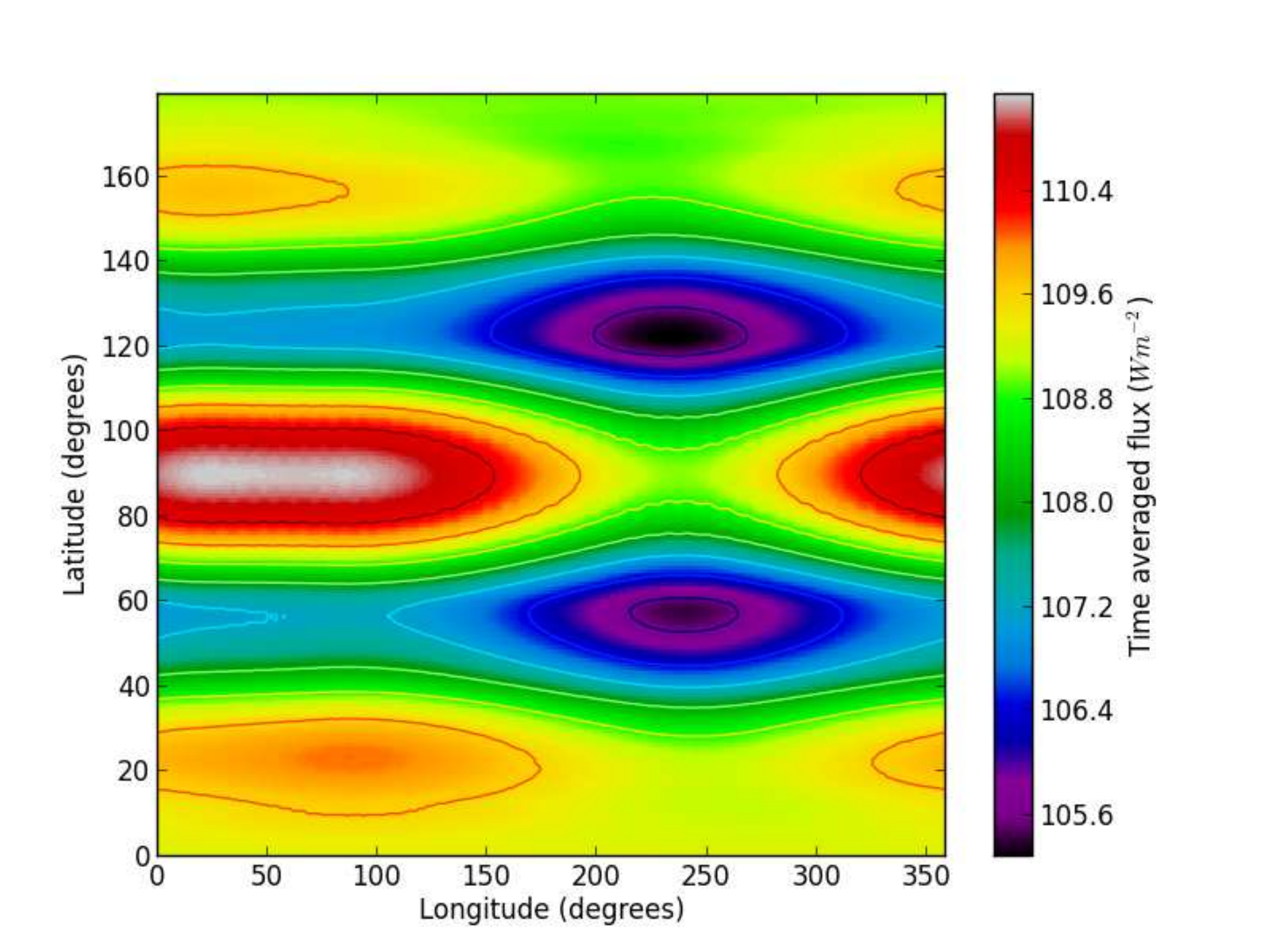} \\
\end{array}$
\caption{Flux patterns (averaged over the planetary orbital period) for a circumbinary planet in orbit around the Kepler-16 binary, with with $a_p=0.7$ AU, and $e_p=0$ (top row), $e_p=0.2$ (middle row) and $e_p=0.5$ (bottom row).  The three columns have obliquities $\delta_p = 0^{\circ},30^{\circ},60^{\circ}$ respectively. \label{fig:flux_kep16}}
\end{figure*}

\noindent This may seem surprising initially, as a back-of-the-envelope calculation would suggest that the eclipse timescale is rather short:

\begin{equation}
t_{eclipse} = \frac{R_1 +R_2}{v_{orbit}} \approx 3 \,\mathrm{hours},
\end{equation}

\noindent where $R_1$ and $R_2$ are the stars' physical radii, and $v_{orbit}$ is the binary's orbital velocity.  However, this equation is valid only for a static observer, and the planet moves at its own orbital velocity relative to the centre of mass of the system.  Therefore the eclipse timescale should be written

\begin{equation}
t_{eclipse} = \frac{R_1 +R_2}{v_{orbit,relative}},
\end{equation}

\begin{figure}
\includegraphics[scale=0.6]{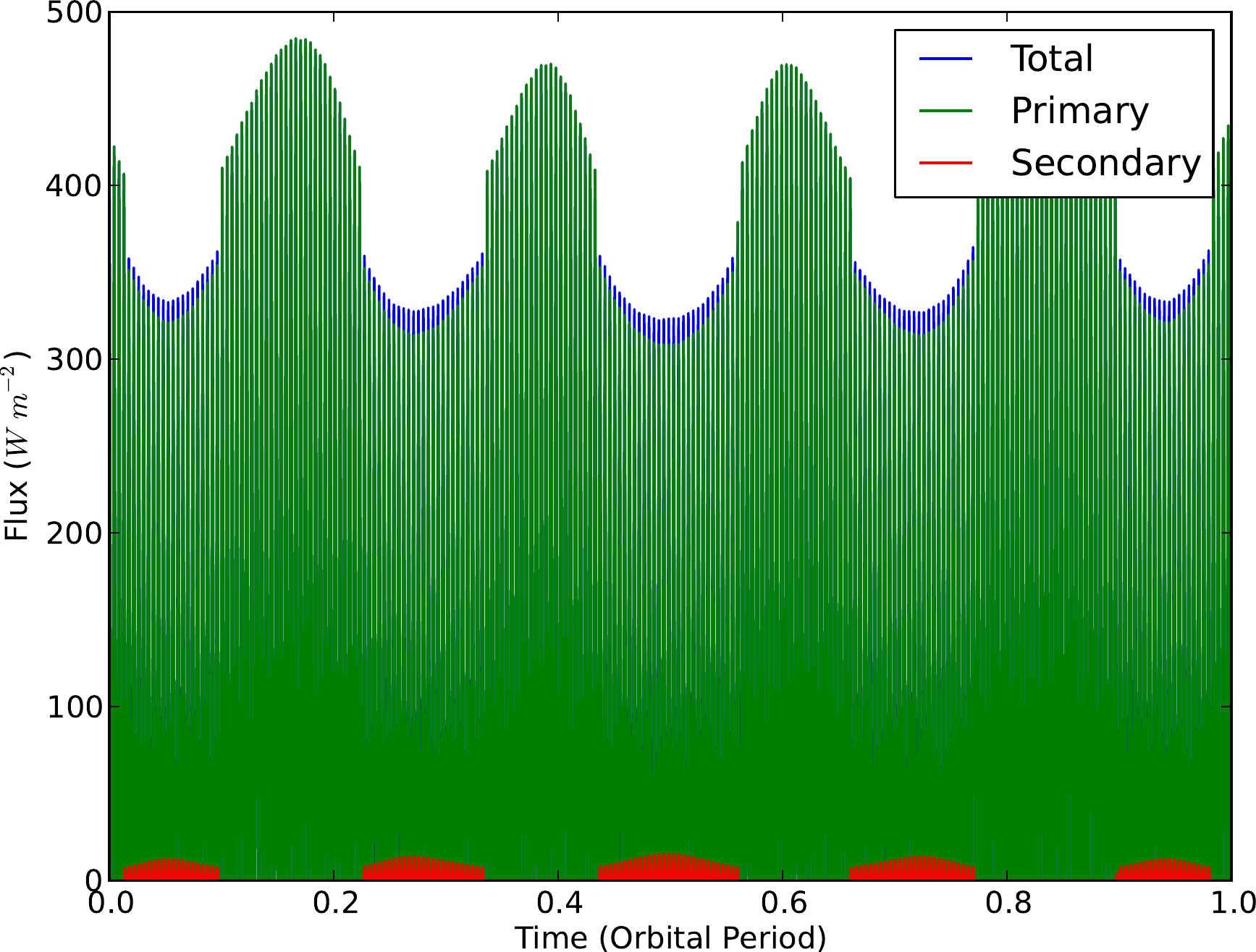}
\caption{The variation of received flux with time, for a planet orbiting the Kepler-16 binary at $a_p=0.7$ AU, with eccentricity $e_p=0$ and obliquity $\delta_p=0^\circ$.  The flux is measured at the planet's equator, along the prime meridian.  Curves are plotted for flux from the primary and secondary, as well as the total received flux. \label{fig:kepler16b_vs_t}}
\end{figure}

\noindent Where $v_{orbit,relative}$ now describes the orbital velocity of the binary in the frame where the planet is at rest. Figure \ref{fig:kepler16b_vs_t} shows how the flux in a single longitude/latitude cell varies with time for a planet orbiting Kepler-16 at $a_p=0.7$ AU, with zero eccentricity and obliquity (the flux is measured at the planet's equator along the prime meridian).

We can see that the total flux varies according to several distinct timescales.  Firstly, there is the standard day/night cycle, which operates at the highest frequency.  As this example has $e_p=0$, $\delta_p=0^\circ$, this frequency is fixed.  The motions of the binary around the centre of mass add an amplitude modulation to this signal, with a period equal to the binary's period.  Finally, the effect of eclipses reduces or completely erases the flux at regular intervals - we can see that the eclipse timescale is indeed much longer than a few hours. As the secondary has a smaller radius than the primary, the primary is only partially eclipsed by the secondary, but the secondary is totally eclipsed by the primary.



\subsubsection{Darkness Patterns}

\noindent For a planet in a circular orbit around a single star, the total amount of time spent in darkness, for any given longitude and latitude, will be half of the orbital period\footnote{In reality, this is never precisely correct, due to atmospheric refraction increasing the length of sunsets}.  If the planet's orbit is eccentric, then the changing orbital velocity will ensure that for non-zero obliquity, the planet's northern hemisphere can experience a longer winter than the southern hemisphere (for example).

\begin{figure*}
$\begin{array}{c}
\includegraphics[scale=0.4]{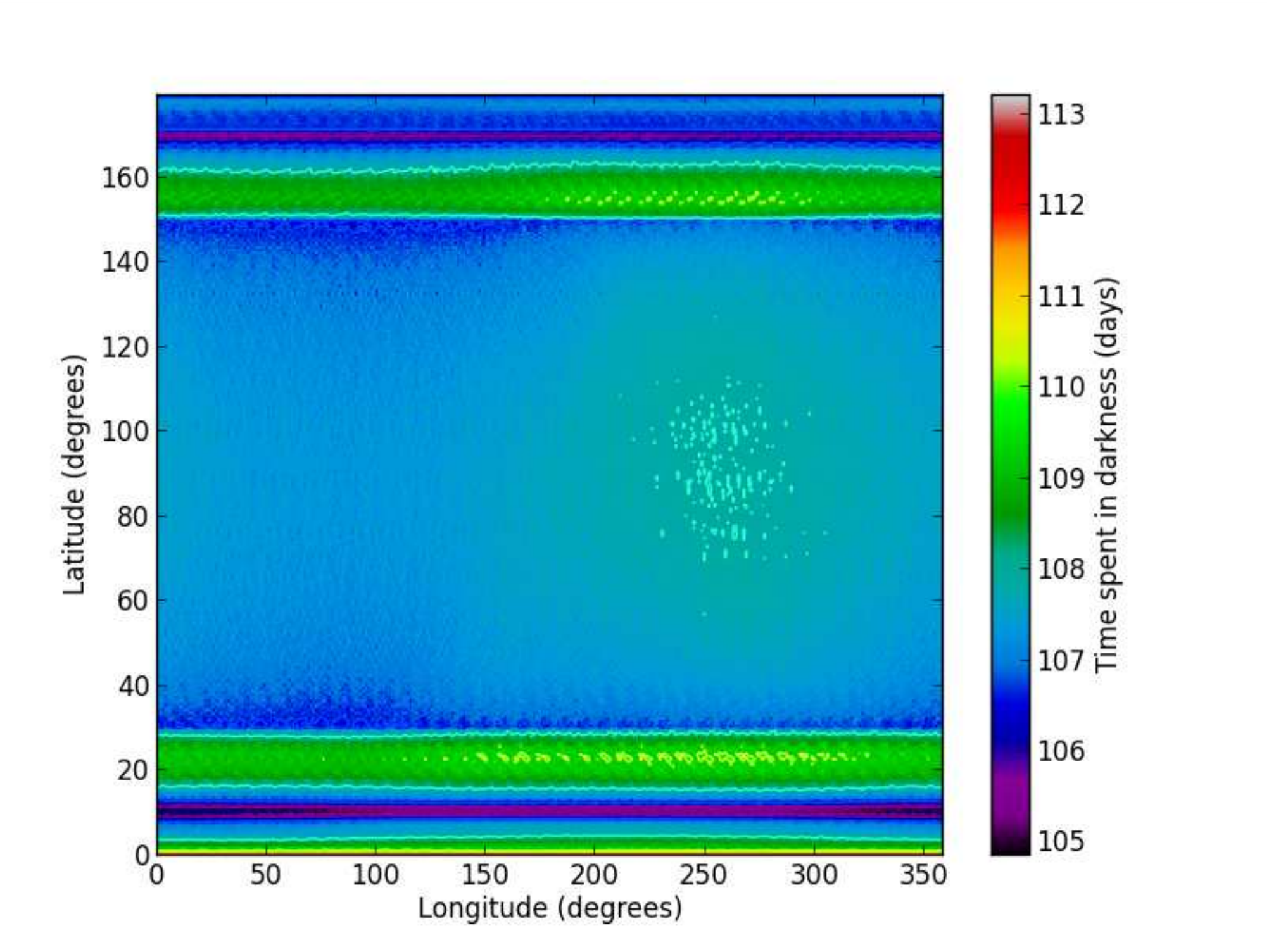} \\
\includegraphics[scale=0.4]{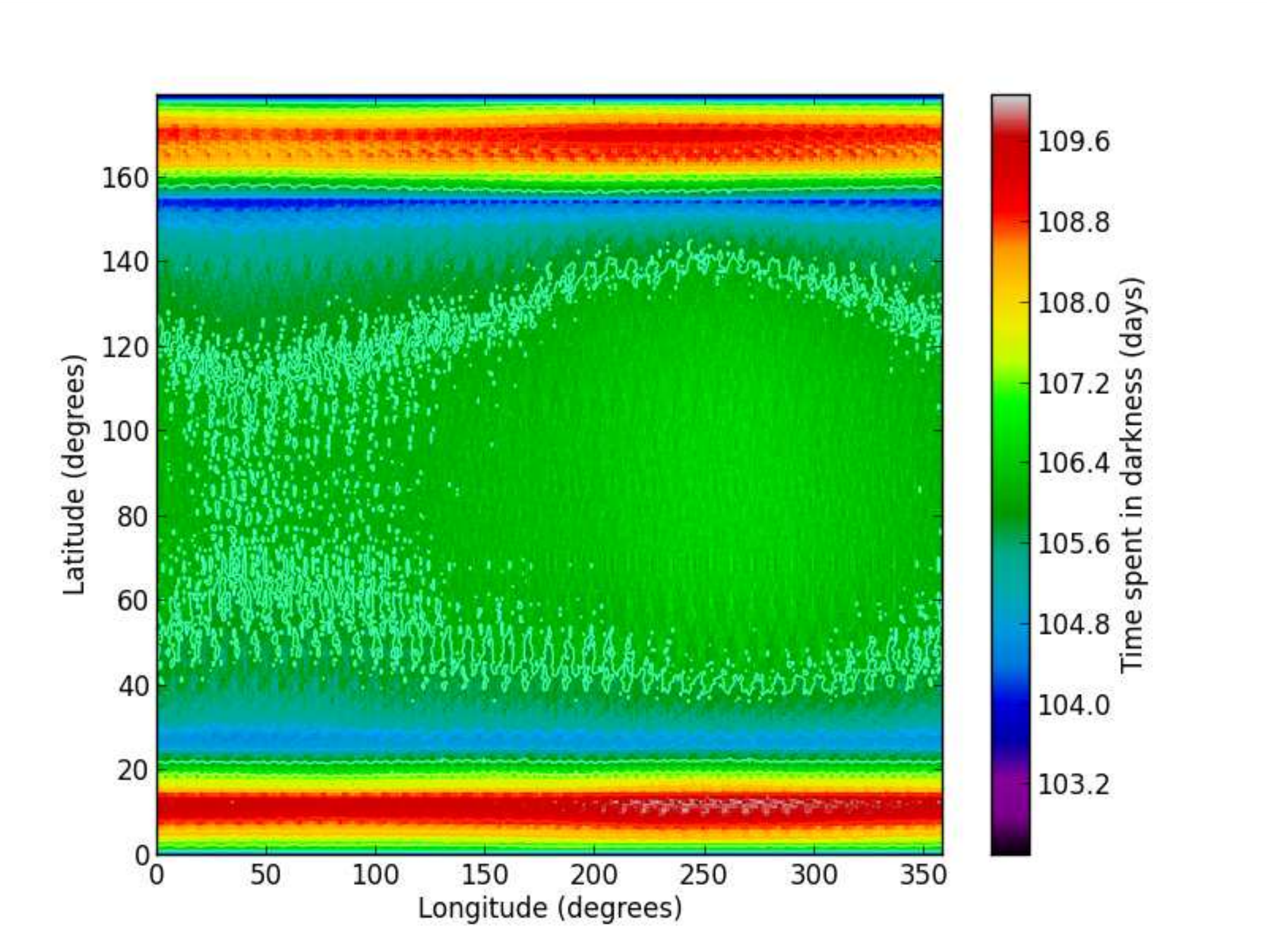} \\
\includegraphics[scale=0.4]{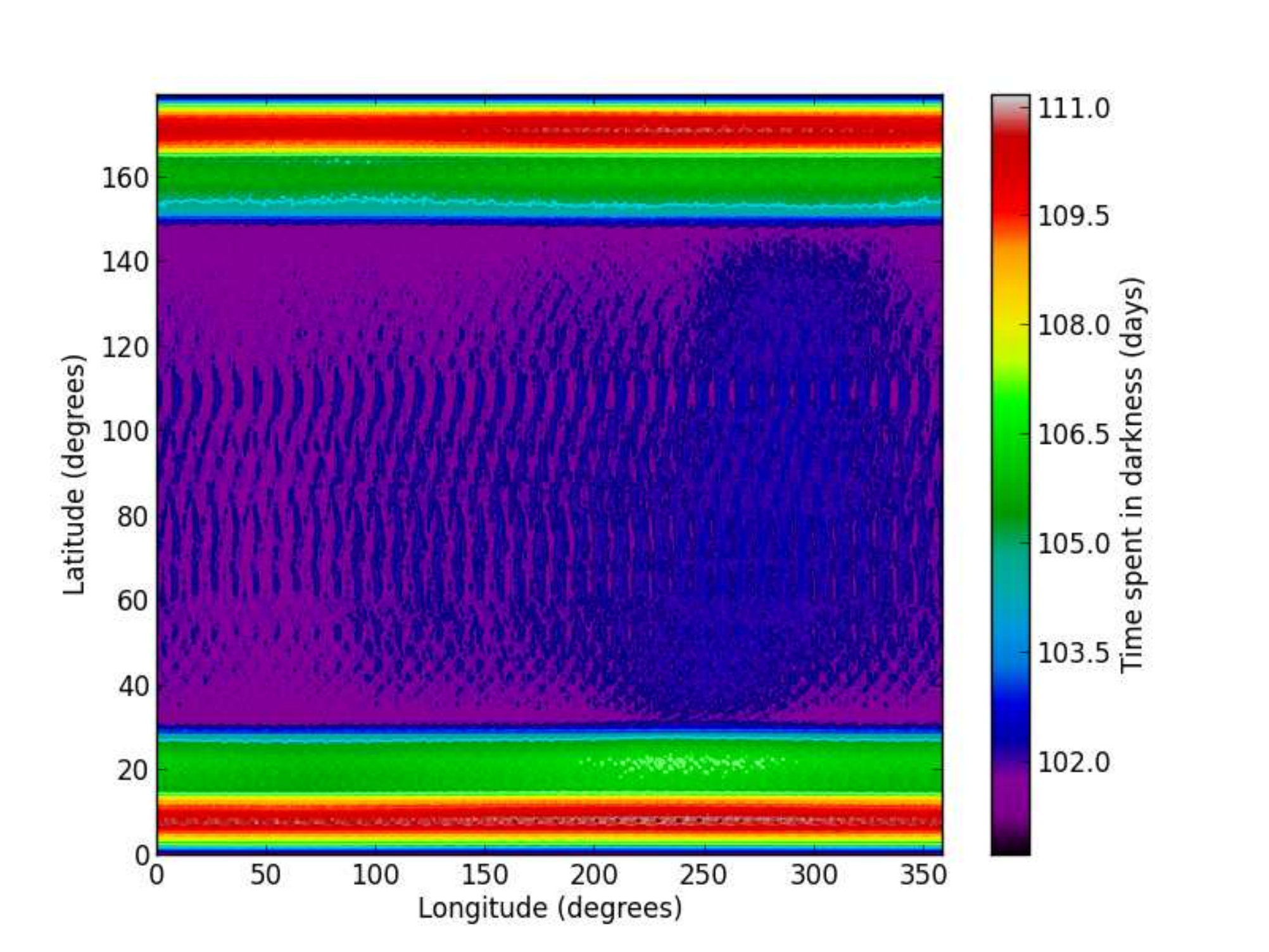} \\
\end{array}$
\caption{Darkness patterns for (left column) a circumbinary planet in orbit around the Kepler-16 binary, each with $a_p=0.7$ AU, obliquity $\delta_p=30^{\circ}$, and (top row) $e_p=0$, (middle row) $e_p=0.2$, (bottom row)  $e_p=0.5$.\label{fig:dark_kep16}}
\end{figure*}

If the planet is in a binary system, then for a given location on its surface to be in darkness, both stars must not be in the sky.  Figure \ref{fig:dark_kep16} shows the total time spent in darkness for planets with obliquity $\delta_p=30^\circ$ orbiting the Kepler-16 binary.  We also run (but do not show here) simulations with the secondary removed, to assess our ability to measure uniform darkness times.  

In these single star runs, for circular orbits, the darkness time in the single star case is quite uniform, to within the numerical limits of the simulation (i.e. for $\Delta t\approx 30\, \mathrm{mins}$).  The range in measured darkness time is a few hours for the circular orbit case (this value is so large due to difficulties measuring the darkness period at the polar circles of the planet).  As the eccentricity increases, it is clear that some longitudes receive less daylight than others, and this is a function of orbital phase.  The limits introduced by a finite timestep introduces problems resolving along the latitude of the polar circles - however, it is clear that for a single star system there is no evidence of any latitudinal dependence.

In the two-star case (Figure \ref{fig:dark_kep16}) the difference is stark.  The overall time spent in darkness is shorter - this is a natural consequence of having a second radiation source.  However, we can see a clear latitudinal differentiation, demarcated by the polar circles.  These are regions where darkness can occur in protracted, uninterrupted intervals as well as part of the night/day cycle.  As the planet orbits the centre of mass of the system, the motion of the stars can be sufficient to hide them from view when the higher latitudes would normally begin their exit from winter, if their radiation source resided at the local centre of mass.  This introduces a slight delay on when daylight returns to these regions, adding extra darkness time.  Even in the case of a circular orbit, the difference in darkness time between the polar and equator regions can be as large as 8 days!

\subsection{Kepler-47c}

\noindent Kepler-47 is composed of a G star and an M star, with masses $M_1=1.043 \msol$ and $M_2=0.362 \msol$ respectively.  Their orbit has semimajor axis $a_{bin}=0.0836$ AU and eccentricity $e_{bin}=0.0234$.  Other properties are taken from \citet{Orosz2012}.  

Kepler-47 is the first circumbinary planetary system with multiple planet detections.  At an orbital semimajor axis of 0.989 AU, with an eccentricity constrained to be less than 0.4, Kepler-47c is in the habitable zone of this system \citep{Kane2013,Haghighipour2013,Forgan2014}, whereas Kepler-47b is too close to the binary (and subsequently too hot).  

Again, Kepler-47c is not an Earthlike planet, but for our purposes we will assume an Earthlike planet exists at the same semimajor axis as Kepler-47c.  This binary system has a relatively large binary mass ratio, with extremely frequent eclipses (the binary orbital period is approximately 7 days).

\subsubsection{Flux Patterns}

\begin{figure*}
$\begin{array}{ccc}
\includegraphics[scale=0.25]{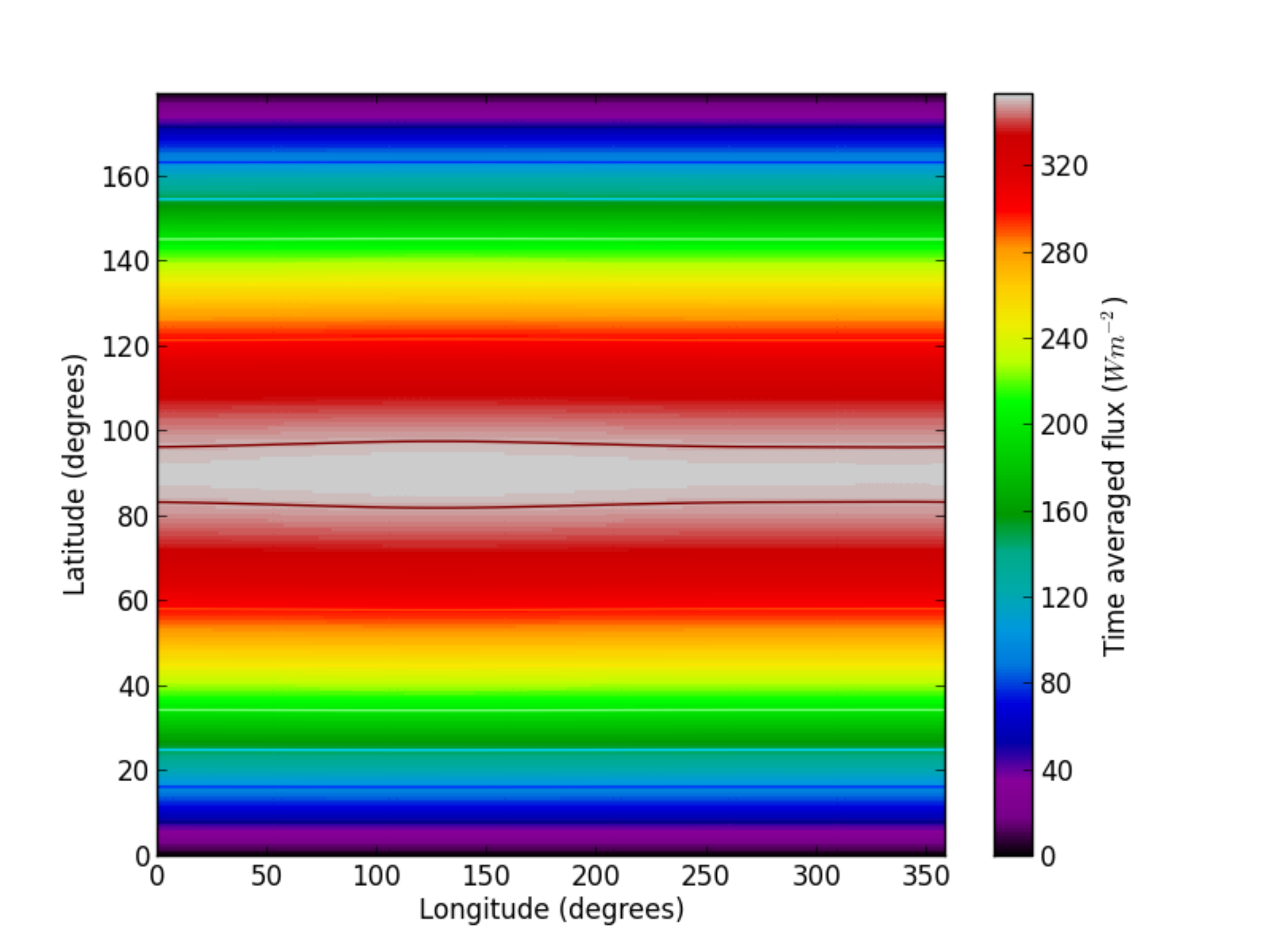} &
\includegraphics[scale=0.25]{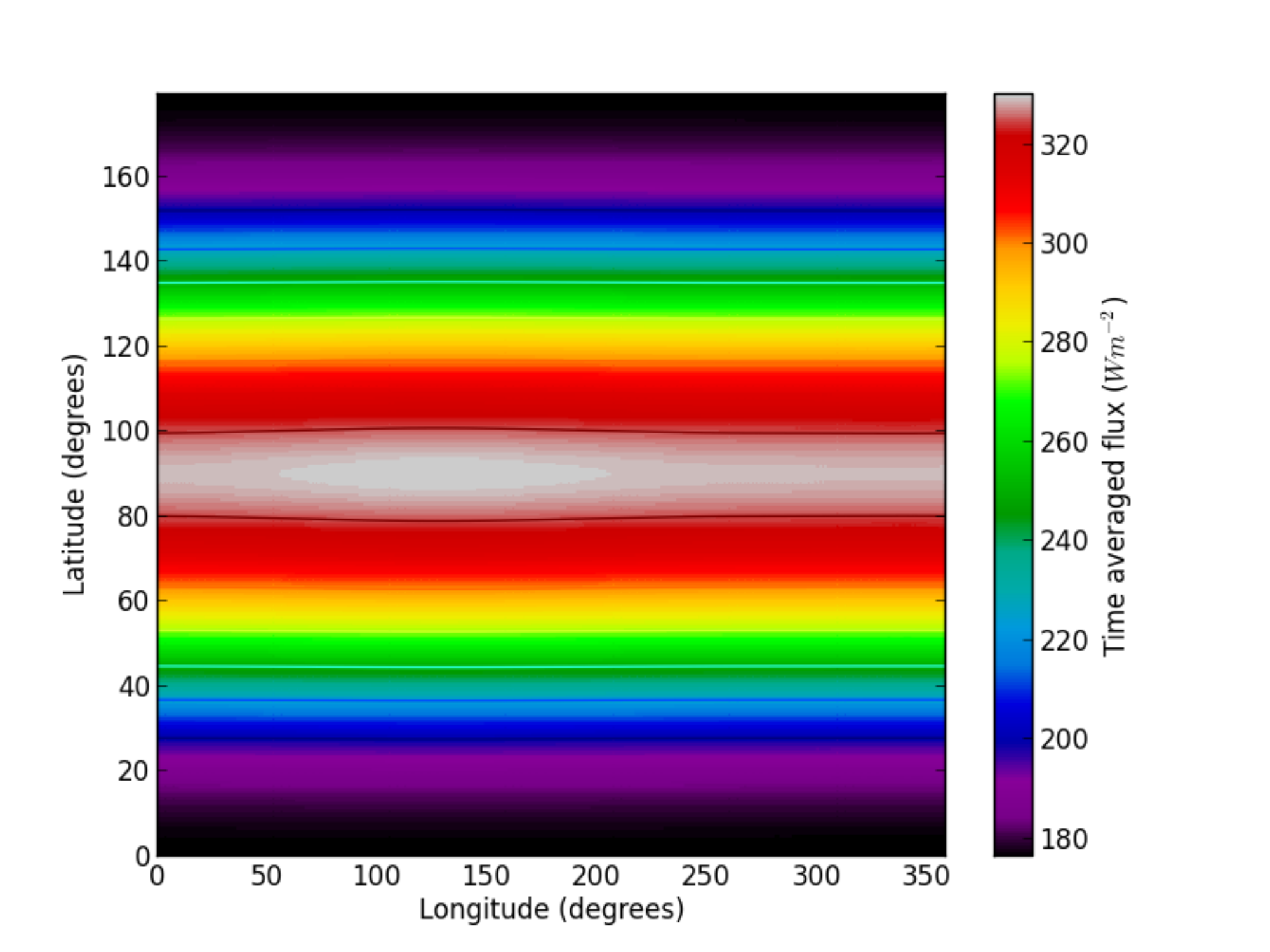} &
\includegraphics[scale=0.25]{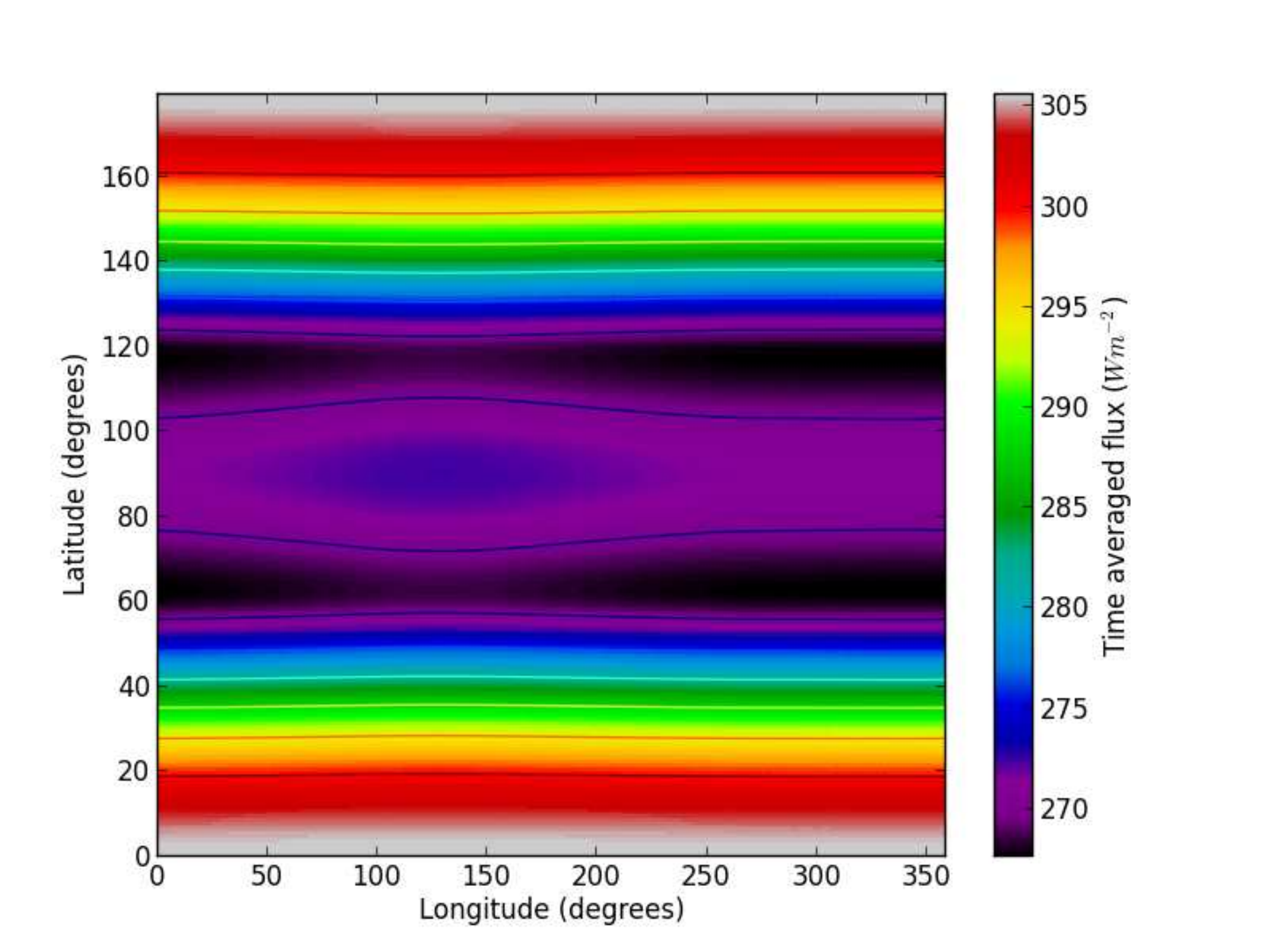} \\
\includegraphics[scale=0.25]{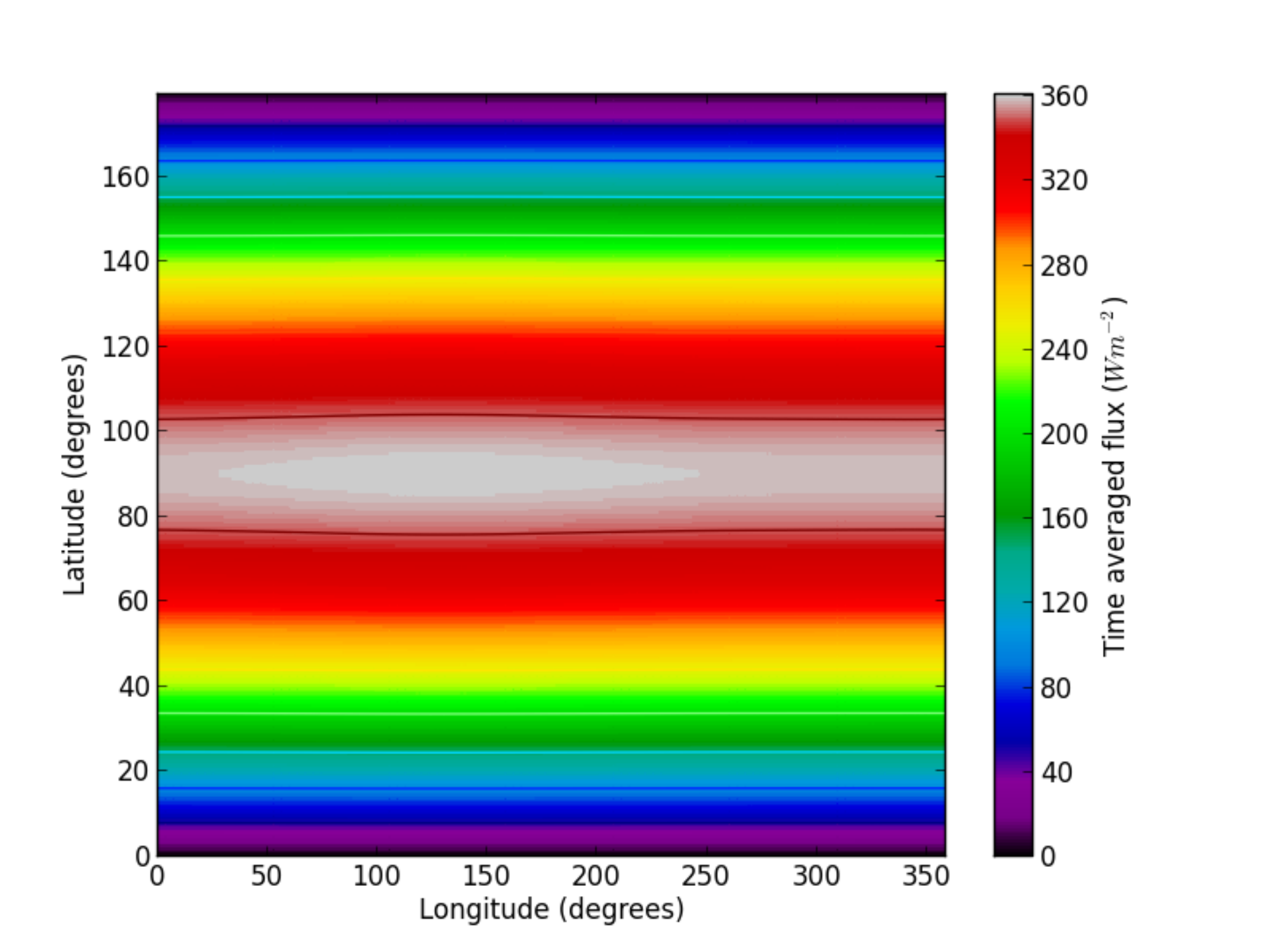} &
\includegraphics[scale=0.25]{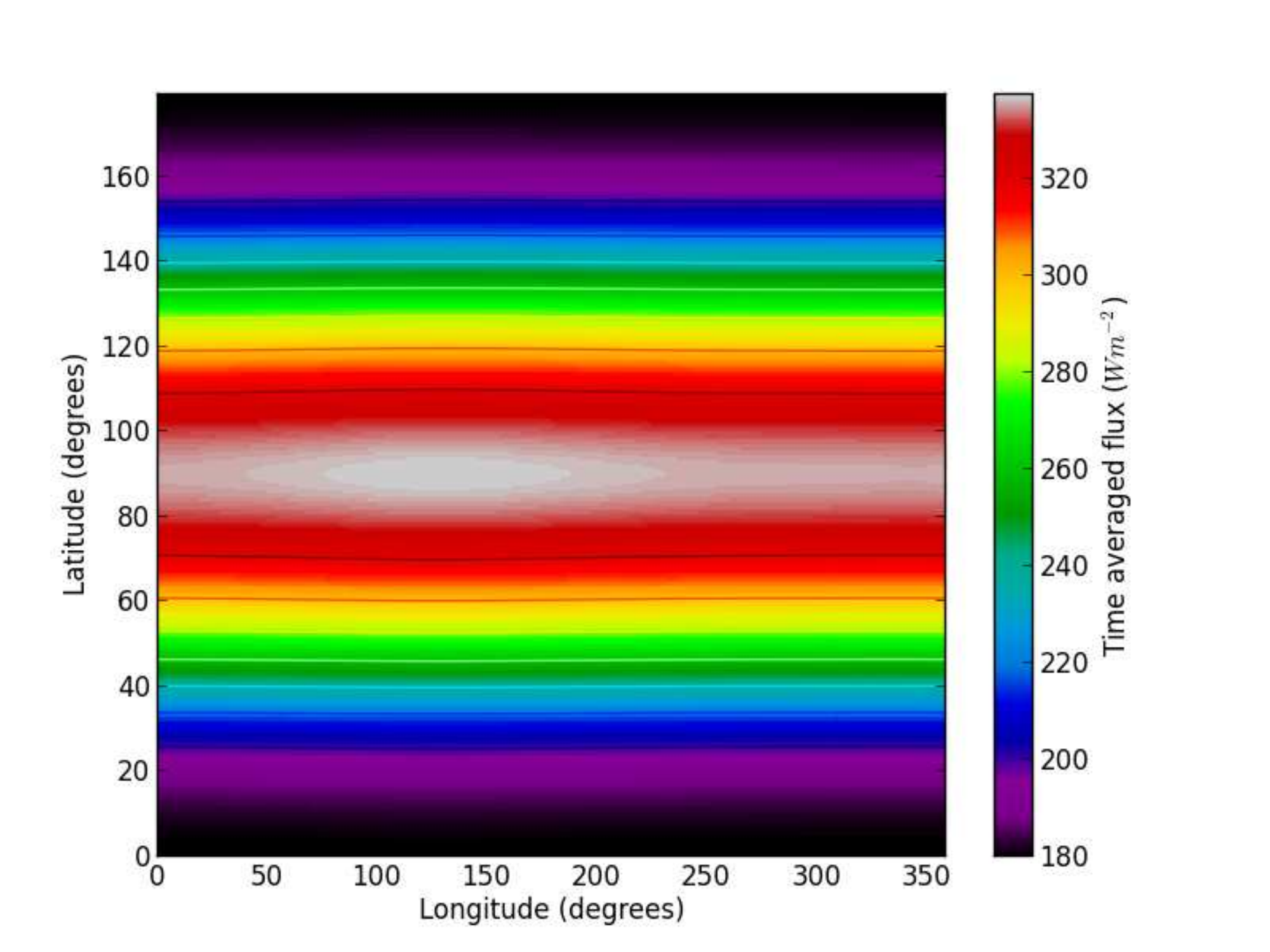} &
\includegraphics[scale=0.25]{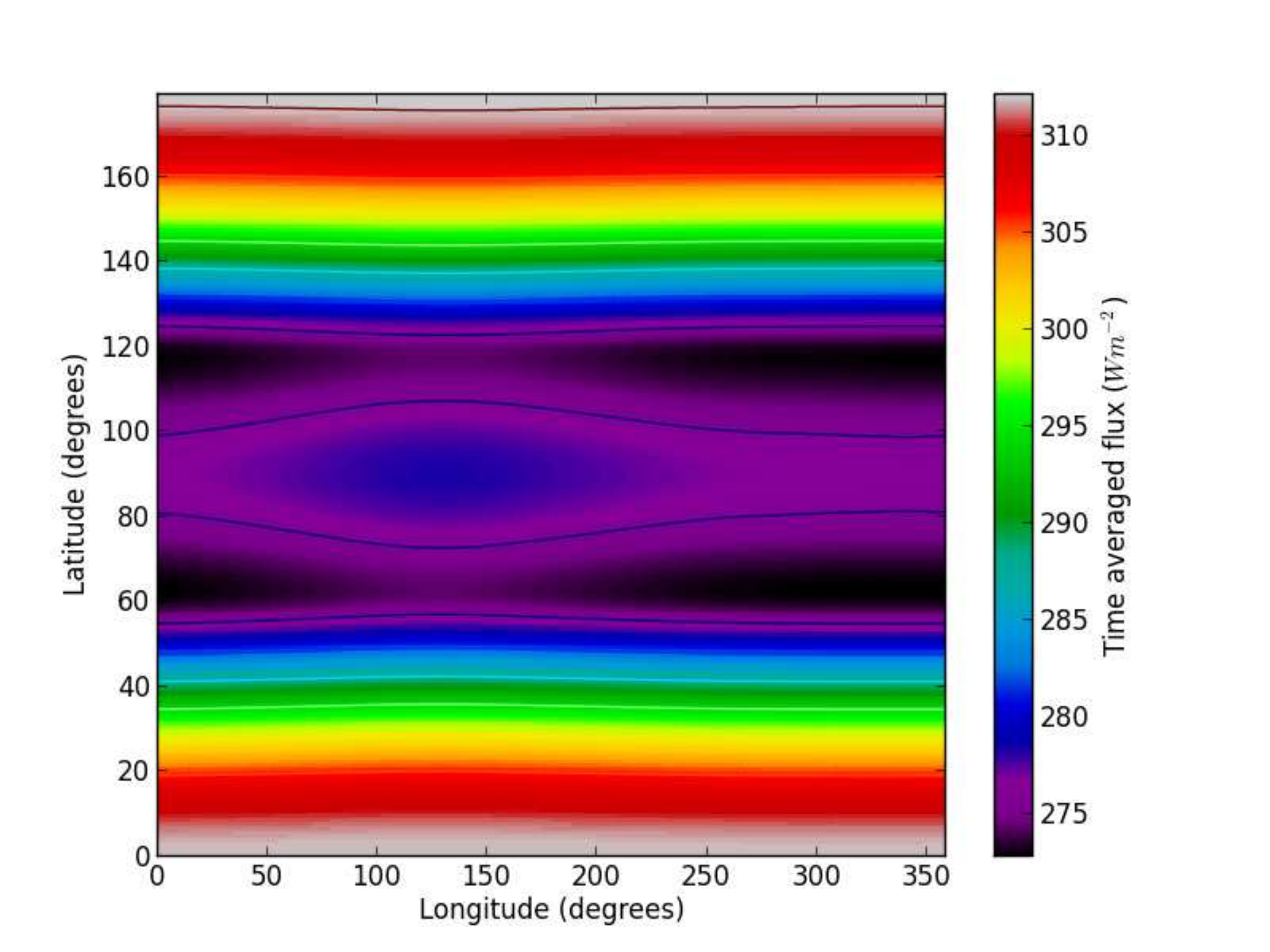} \\
\includegraphics[scale=0.25]{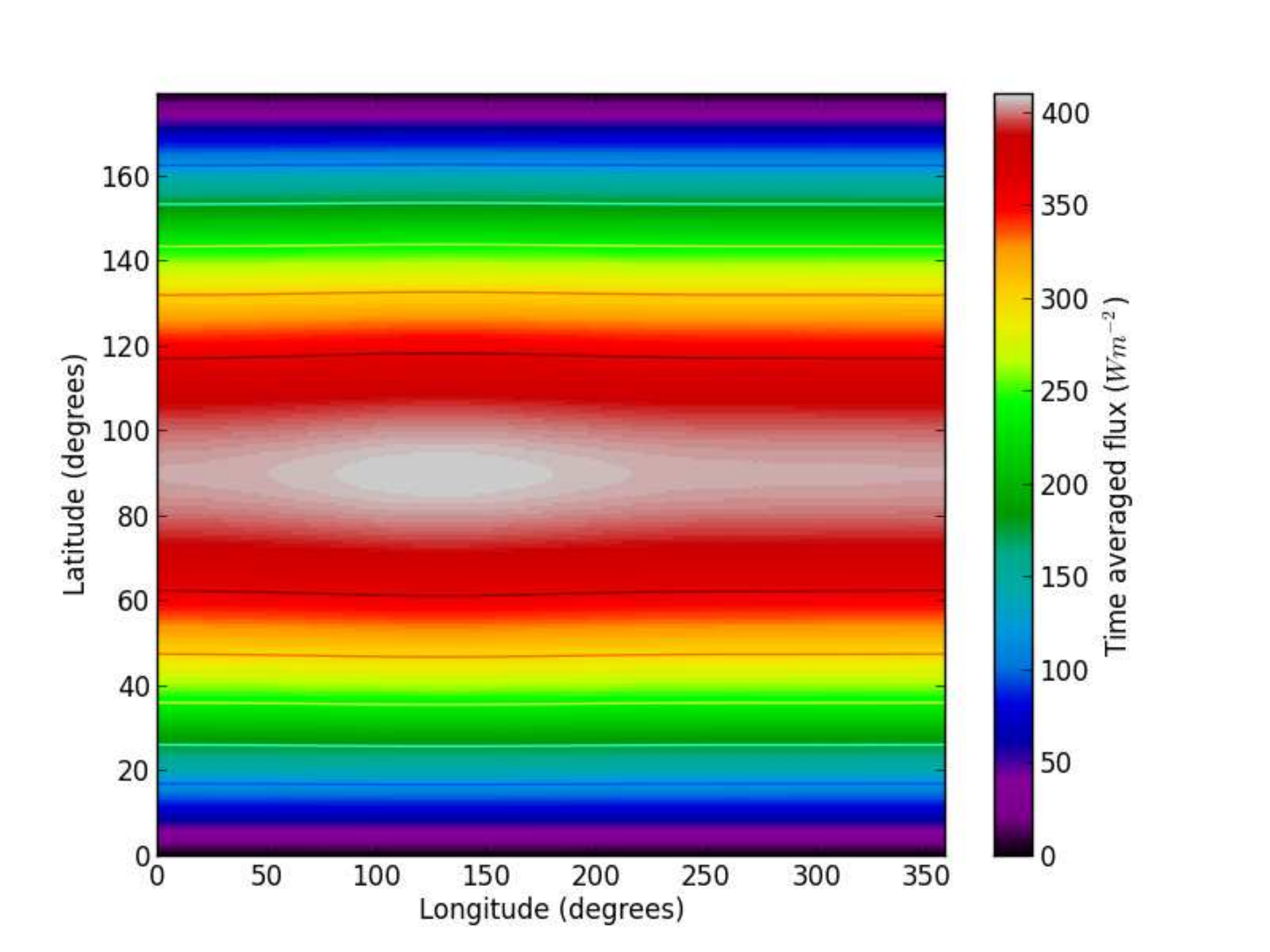} &
\includegraphics[scale=0.25]{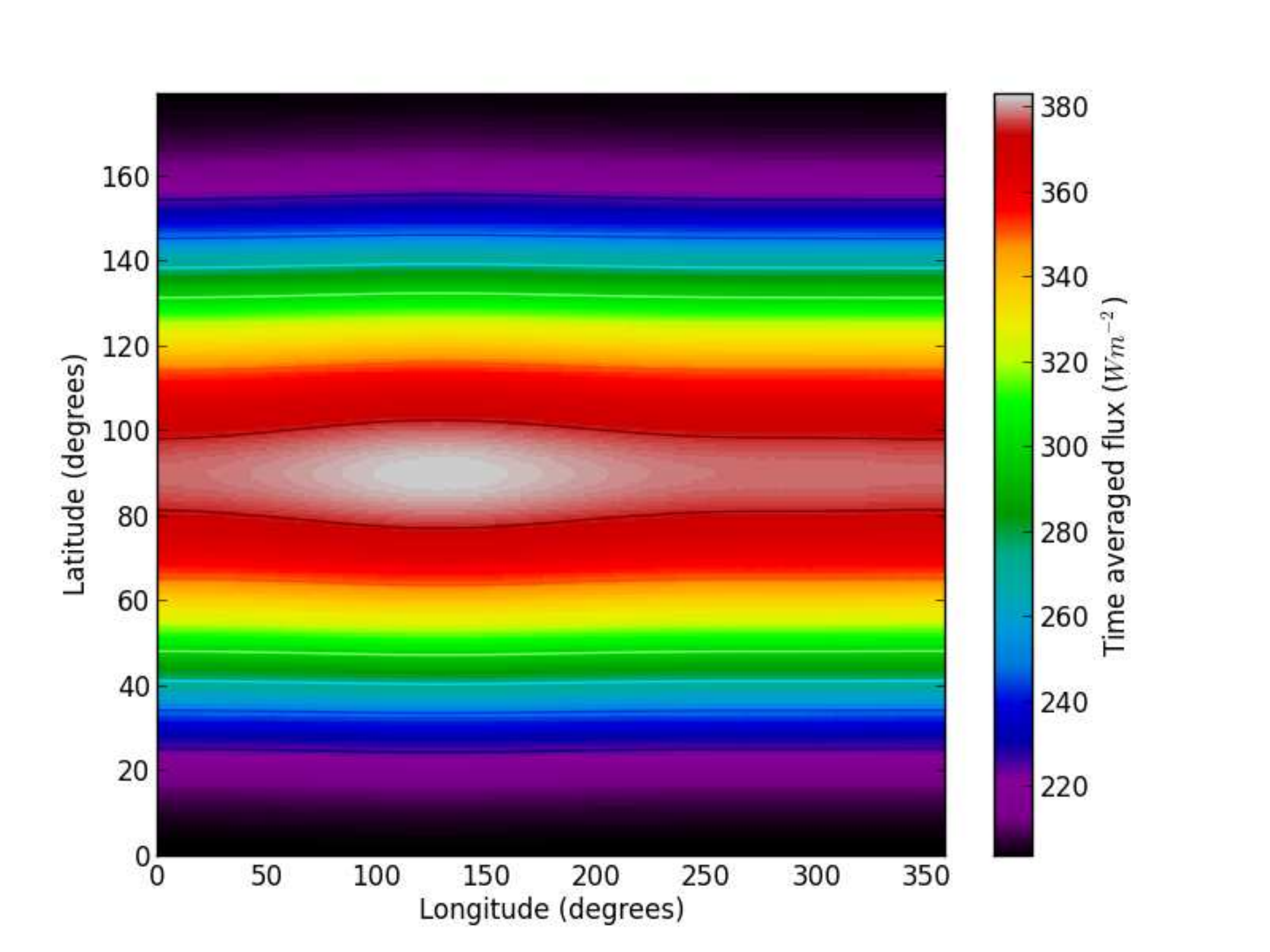} &
\includegraphics[scale=0.25]{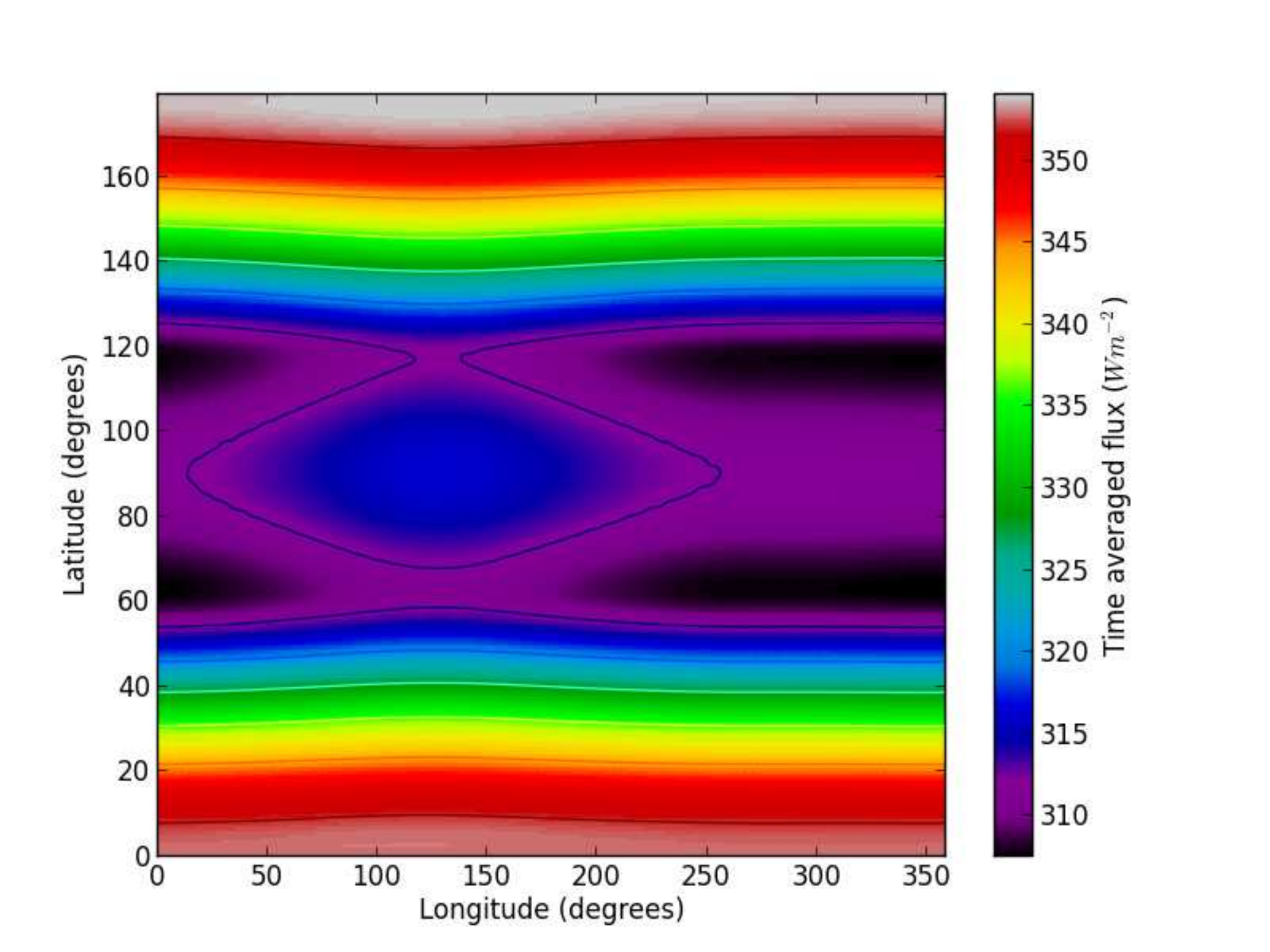} \\
\end{array}$
\caption{Flux patterns (averaged over the planetary orbital period) for a circumbinary planet in orbit around the Kepler-47 binary, with $a_p=0.989$ AU, and $e_p=0$ (top row), $e_p=0.2$ (middle row) and $e_p=0.5$ (bottom row).  The three columns have obliquities $\delta_p = 0^{\circ},30^{\circ},60^{\circ}$ respectively. \label{fig:flux_kep47}}
\end{figure*}

\noindent Figure \ref{fig:flux_kep47} shows the time integrated flux for planets orbiting the Kepler-47c binary at $a_p=0.989$ AU.  We again vary the eccentricity between $e_p = 0,0.2,0.5$ (top, middle, bottom rows), and the obliquity between $\delta_p = 0^{\circ},30^{\circ},60^{\circ}$ (left, middle, right columns).

The low obliquity cases are very similar to those for Kepler-16b.  Again, the peak flux is spread over a wider range of longitudes than would be seen for a single star - however, the relatively small $a_{bin}$ places the two substellar points closer together on the planet's surface, and so the smearing of flux is less evident.  Also, the secondary luminosity is a good deal smaller than the primary luminosity, so it contributes much less to the total surface flux, relatively speaking.

\begin{figure}
\includegraphics[scale=0.6]{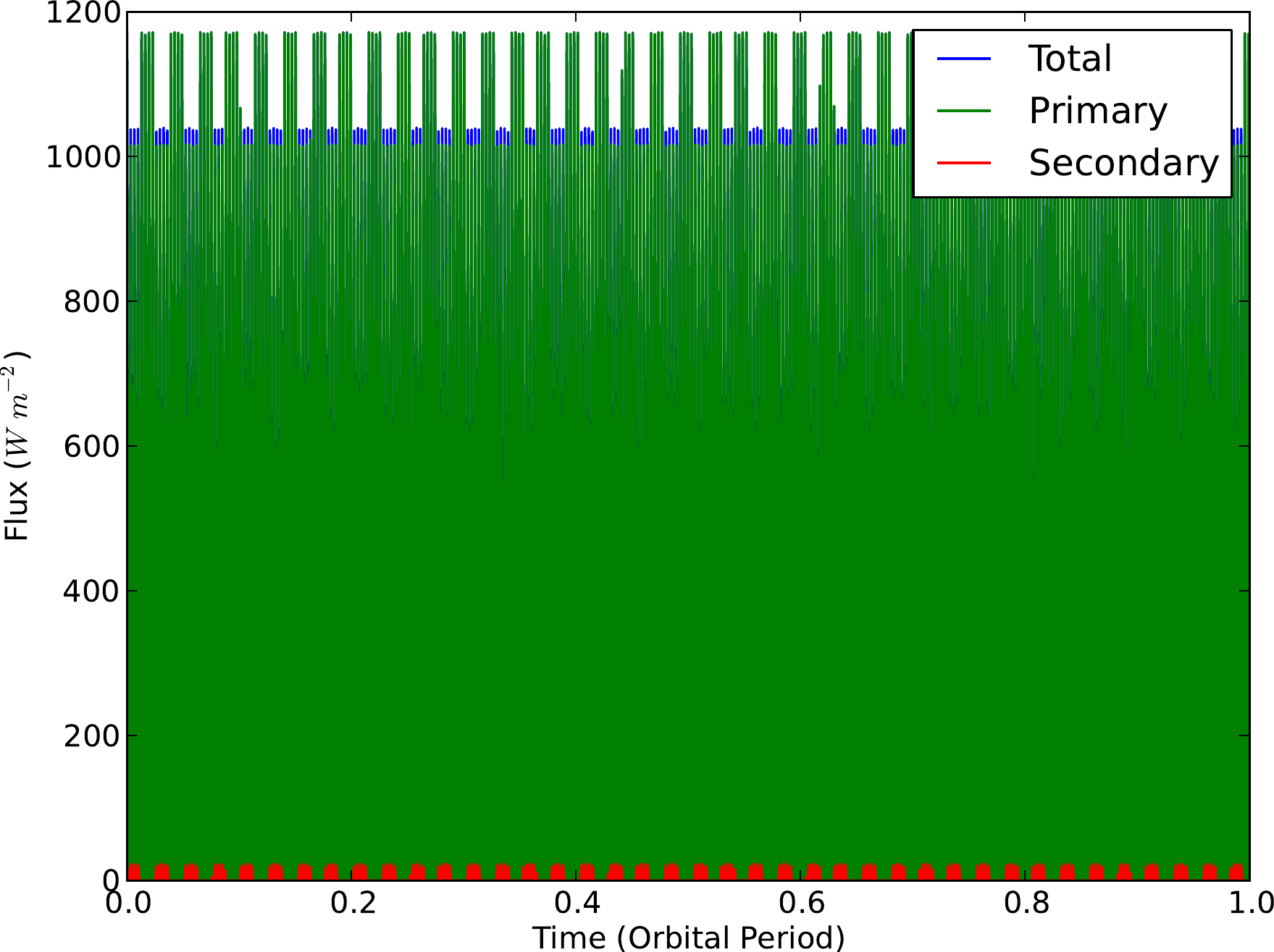}
\caption{The variation of received flux with time, for a planet orbiting the Kepler-47 binary at $a_p=0.989$ AU, with eccentricity $e_p=0$ and obliquity $\delta_p=0^\circ$.  The flux is measured at the planet's equator, along the prime meridian.  Curves are plotted for flux from the primary and secondary, as well as the total received flux. \label{fig:kepler47c_vs_t}}
\end{figure}

For high obliquities, we do not see the poles receiving significantly different levels of flux.  The eclipse timescale is significantly smaller than that of Kepler-16b - as a result, for circular orbits the planet will experience around 40 primary eclipse events per orbit (see Figure  \ref{fig:kepler47c_vs_t}).  Consequently, both the northern and southern hemispheres will experience very similar numbers of eclipses during their summer periods.  Also, the large mass ratio of the binary ensures that eclipses of the secondary by the primary have an almost negligible effect on the incoming flux.

\noindent Figure \ref{fig:kepler47c_vs_t} demonstrates the minimal effect of the second star on the flux received at a given latitude/longitude.  Eclipses of the primary cause a drop in flux of around 15\%, but the event is brief, lasting approximately 3 hours.  

\subsubsection{Darkness Patterns}

\begin{figure*}
$\begin{array}{cc}
\includegraphics[scale=0.4]{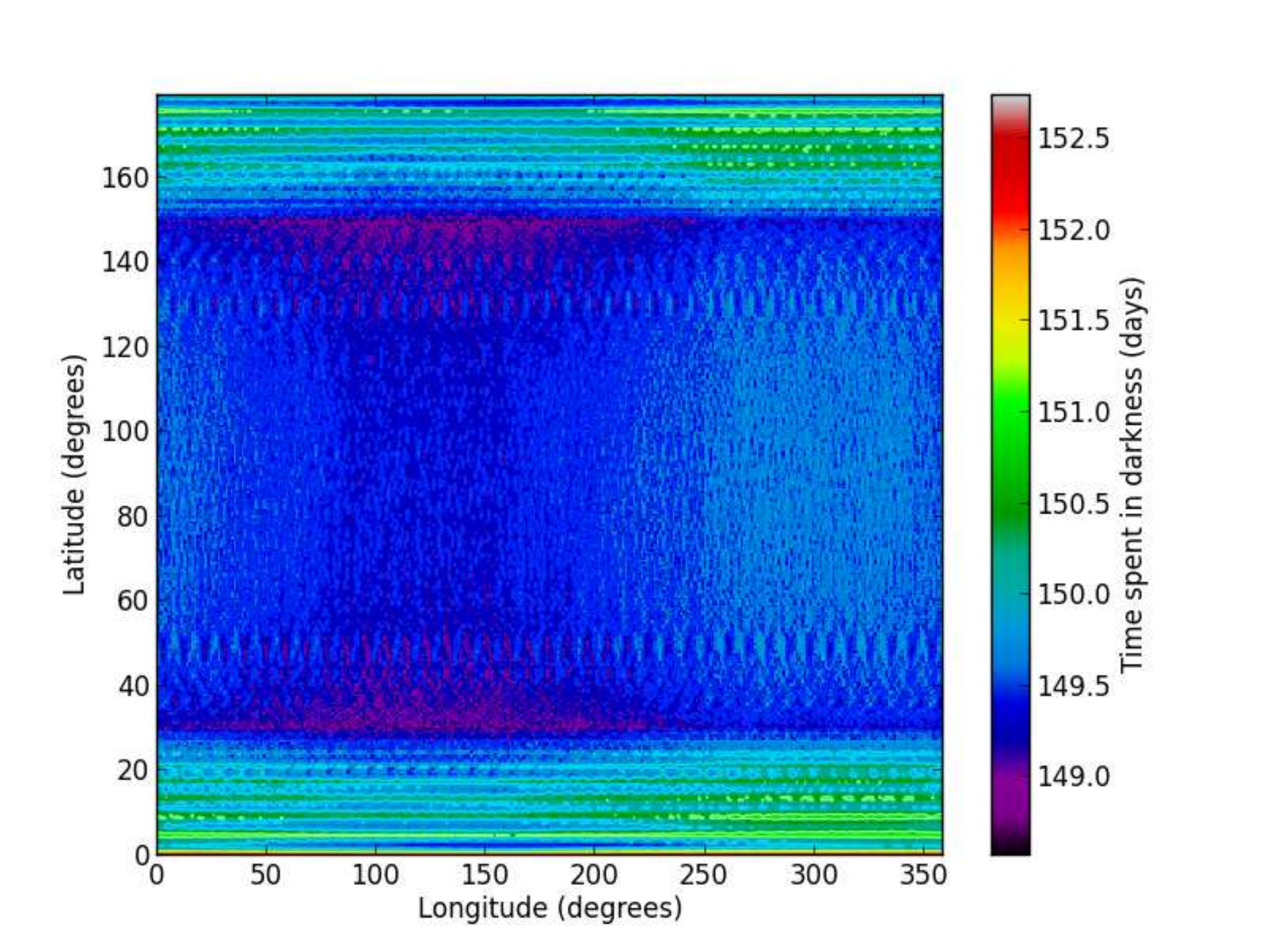} \\
\includegraphics[scale=0.4]{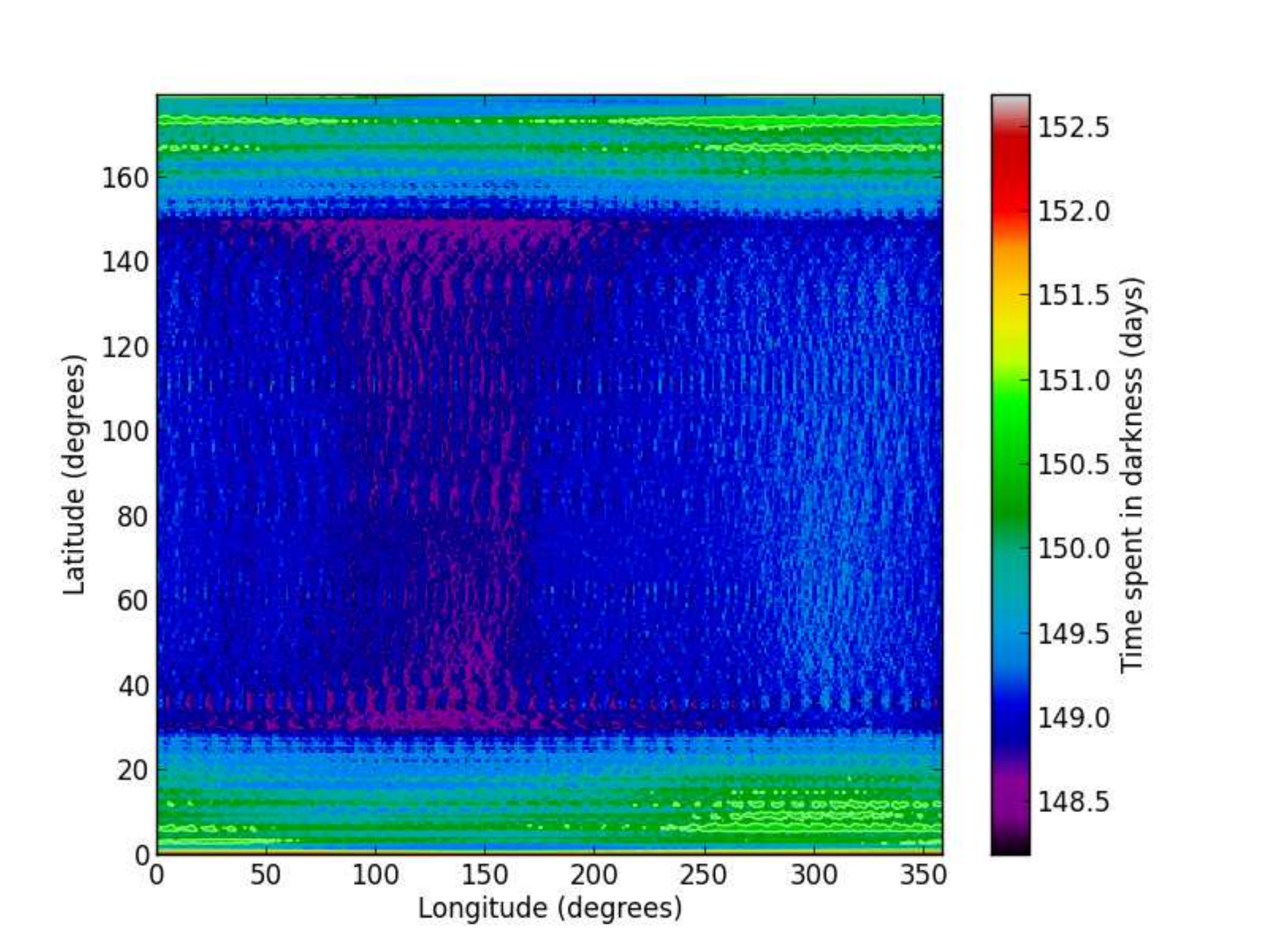} \\
\includegraphics[scale=0.4]{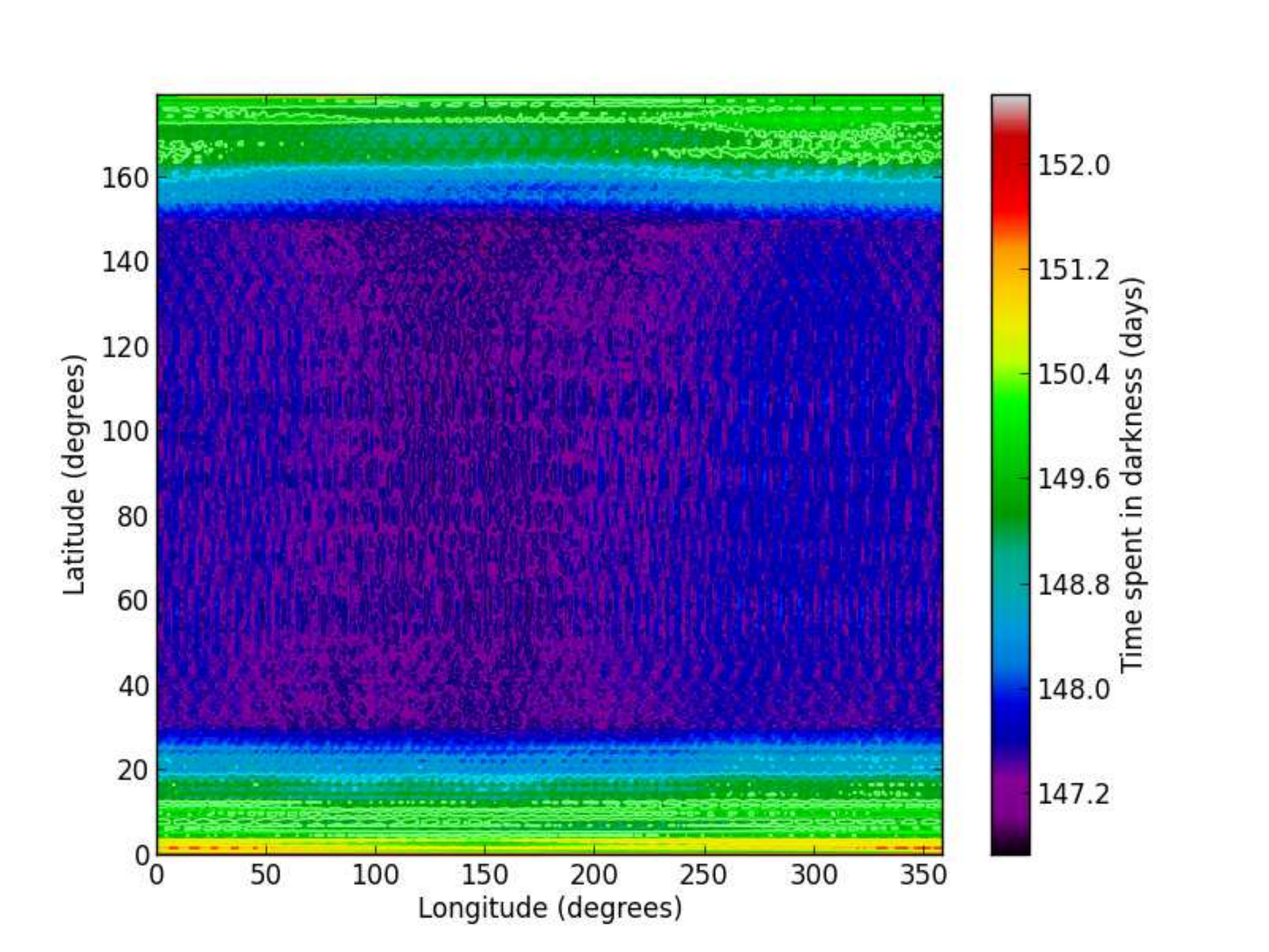} \\
\end{array}$
\caption{Darkness patterns for a circumbinary planet in orbit around the Kepler-47 binary, each with $a_p=0.989$ AU, obliquity $\delta_p=30^{\circ}$, and (top row) $e_p=0$, (middle row) $e_p=0.2$, (bottom row)  $e_p=0.5$. \label{fig:dark_kep47}}
\end{figure*}

\noindent Figure \ref{fig:dark_kep47} shows the time spent in darkness for a planet with obliquity $\delta_p=30^\circ$, orbiting the Kepler-47 binary.  Again, we see that the single star case (not shown here) has relatively uniform darkness time.  The simulation timestep is kept the same as for Kepler-16, so with Kepler-47c's increased orbital period, the simulation can carry out significantly more timesteps, improving its ability to measure uniform darkness maps.

A circular, zero obliquity orbit around the Kepler-47 binary has a range in darkness time of around 4 days, increasing to around 5 days as the eccentricity is increased.  Again, different darkness regimes are separated by the polar circles.  This is for the same reasons as seen for Kepler-16 - during prolonged darkness periods in the winter, the orbital motions of the stars introduces a slight delay into the first sunrise of the spring.

We see a slight longitudinal dependence appearing as $e_p$ is increased: comparing $\Lambda=120^\circ$ to $\Lambda=300^\circ$ shows a difference in darkness time of around 1 day.  This is two to three times larger than the range seen in single star cases where we expect uniform darkness maps, so we are confident that this effect is not merely numerical.

\section{Discussion}\label{sec:discussion}

\subsection{The effect of orbital inclination}

\begin{figure*}
$\begin{array}{cc}
\includegraphics[scale=0.4]{kepler47c_e0_ob30.pdf} &
\includegraphics[scale=0.4]{kepler47c_dark_e0_ob30.pdf} \\
\includegraphics[scale=0.4]{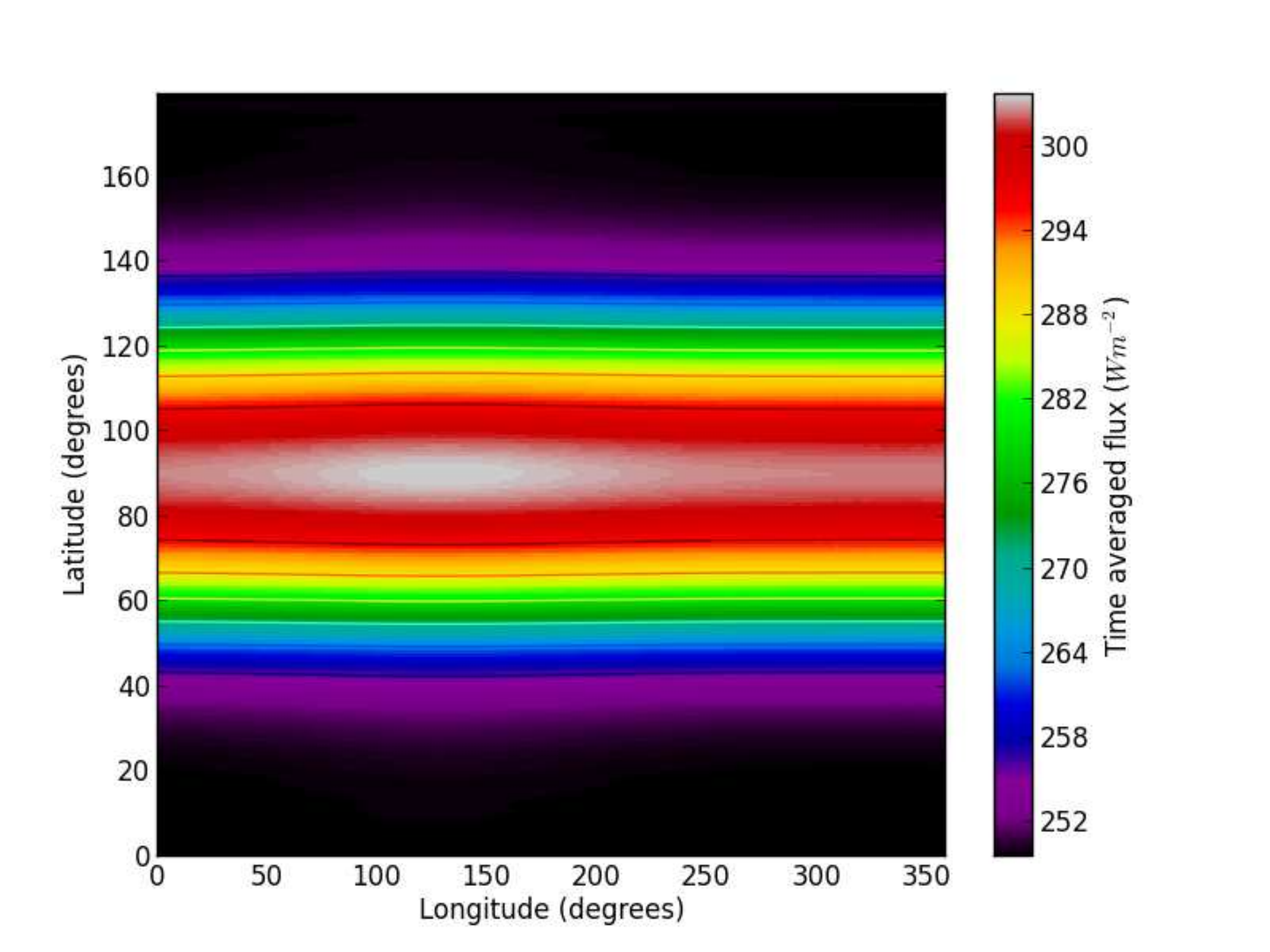} &
\includegraphics[scale=0.4]{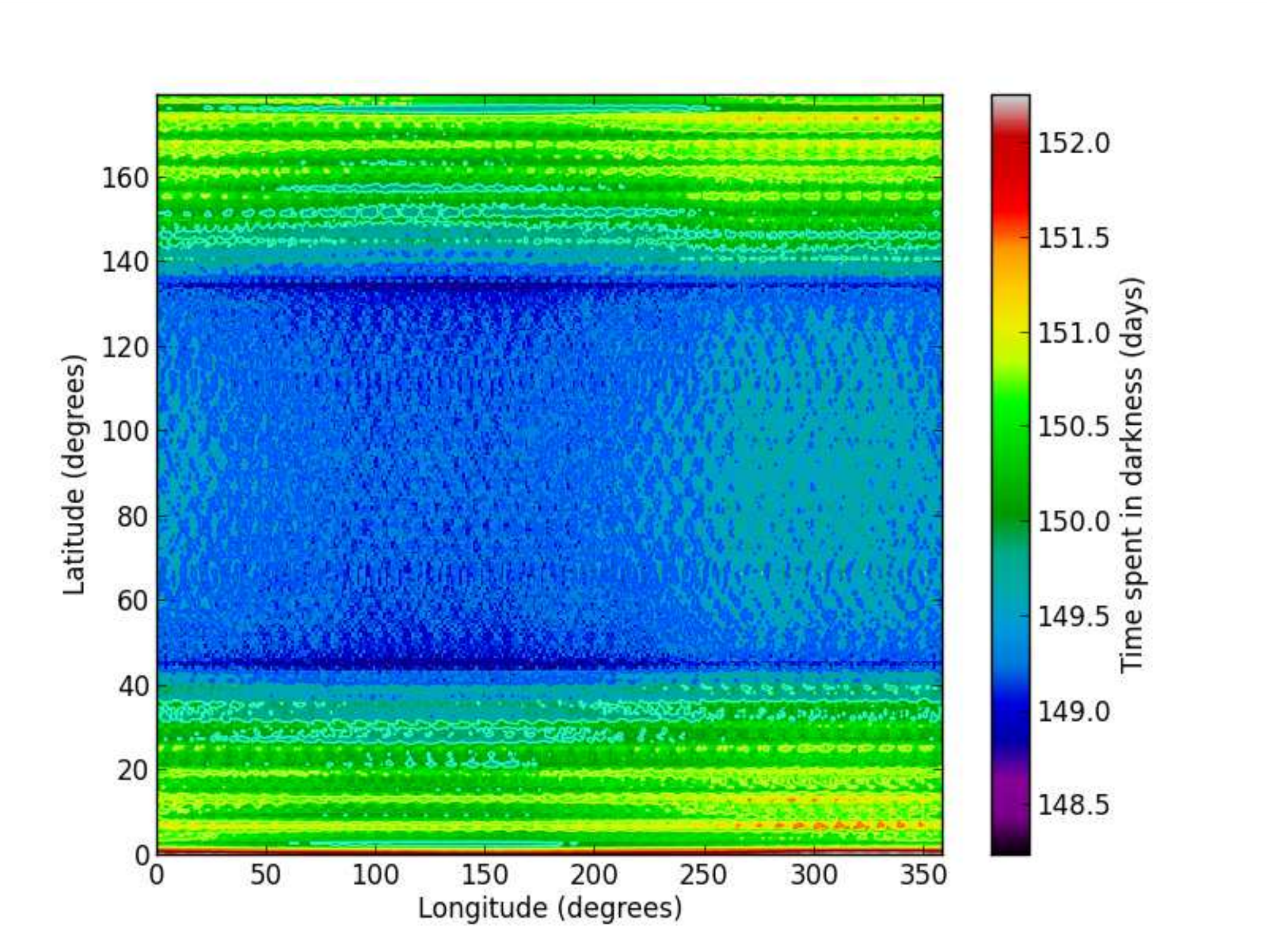} \\
\includegraphics[scale=0.4]{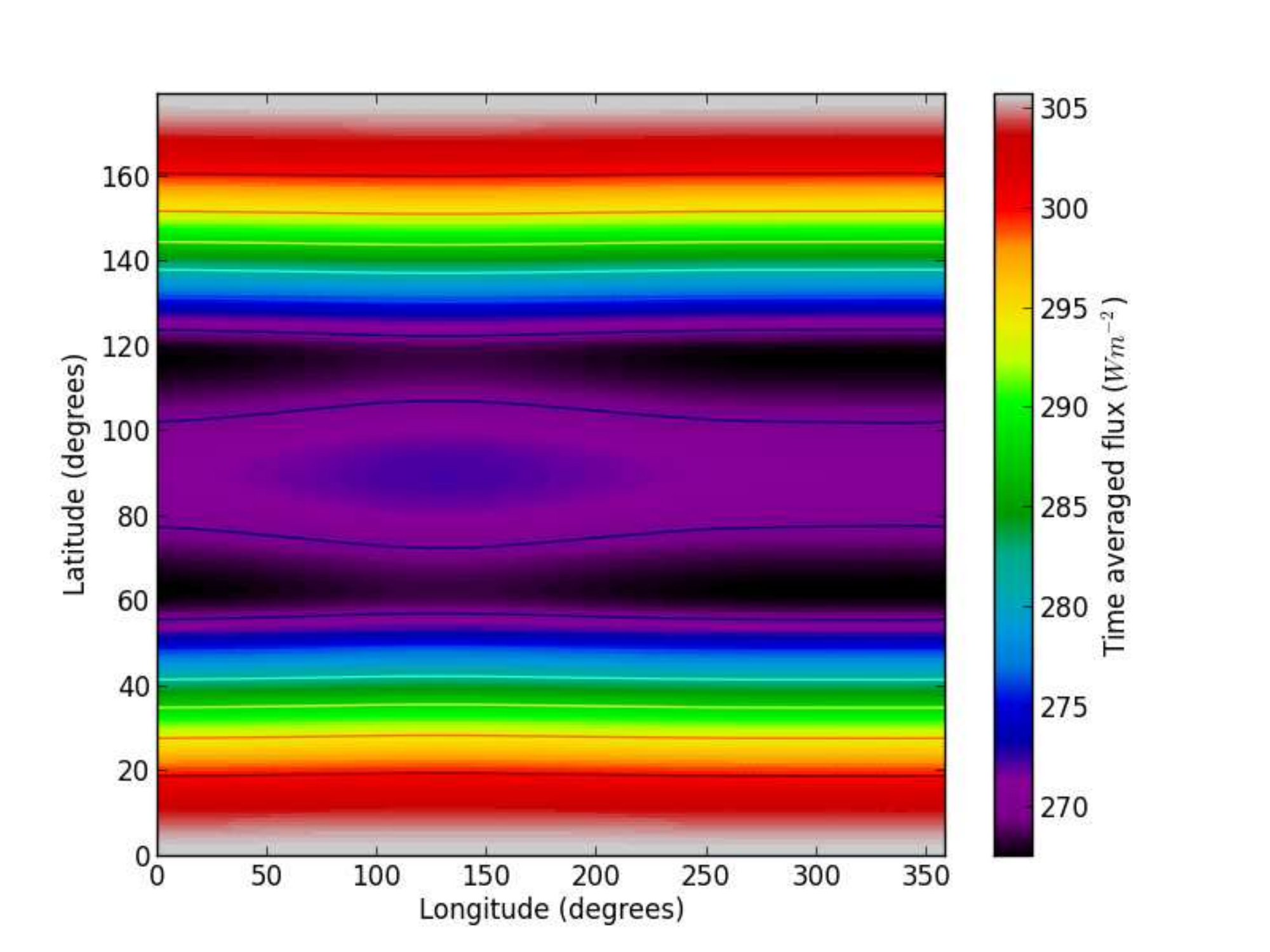} &
\includegraphics[scale=0.4]{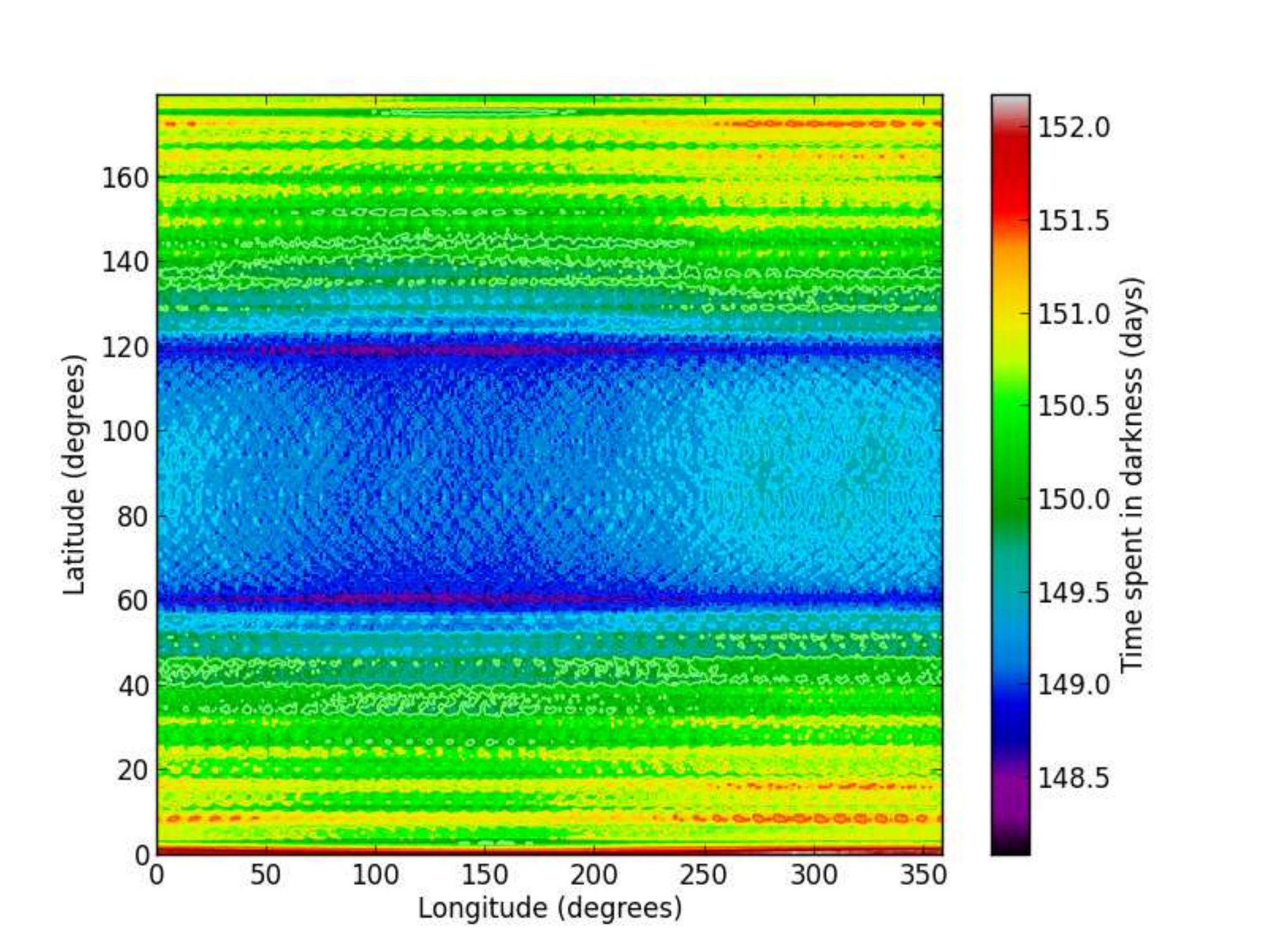} \\
\end{array}$
\caption{The effect of increasing orbital inclination.  Flux patterns (left column) and darkness patterns (right column)for a circumbinary planet in orbit around the Kepler-47 binary with obliquity $\delta_p=30^\circ$, eccentricity $e_p=0$, and (top row) orbital inclination $i_p=0$, (middle row) $i_p=15^\circ$, and (bottom row) $i_p=30^\circ$ \label{fig:inc_kep47}}
\end{figure*}

Figure \ref{fig:inc_kep47} shows dependence of both flux and darkness patterns on orbital inclination for the Kepler-47 system.  As $a_{bin}$ and $e_{bin}$ are both relatively low, the binary sufficiently resembles an extended single star that increasing $i_p$ is equivalent to increasing $\delta_p$, making the flux and darkness patterns difficult to distinguish from a simulation with high obliquity orbiting in the binary plane ($i_p=0$).

\begin{figure*}
$\begin{array}{cc}
\includegraphics[scale=0.4]{kepler16b_e0_ob30.pdf} &
\includegraphics[scale=0.4]{kepler16b_dark_e0_ob30.pdf} \\
\includegraphics[scale=0.4]{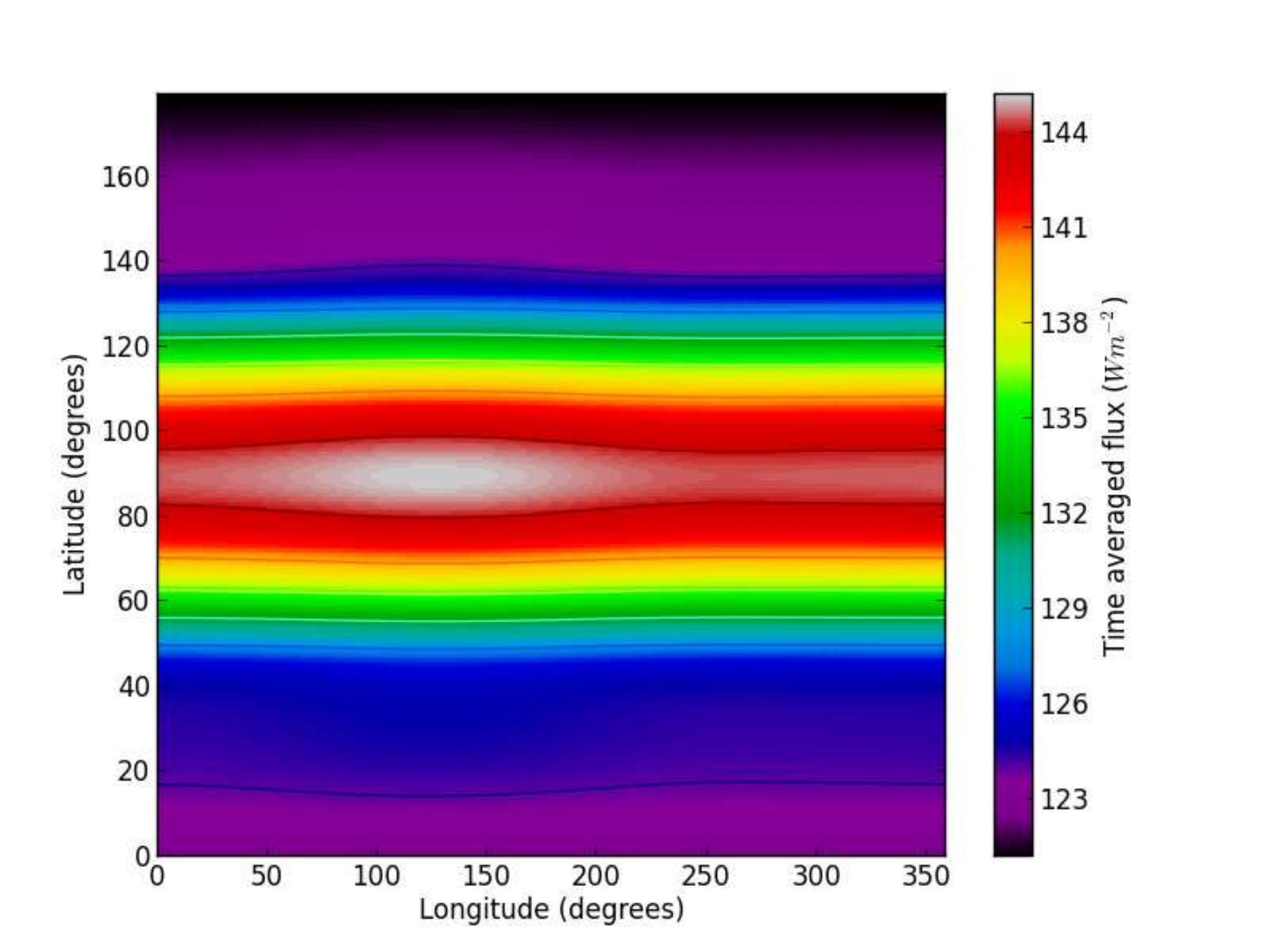} &
\includegraphics[scale=0.4]{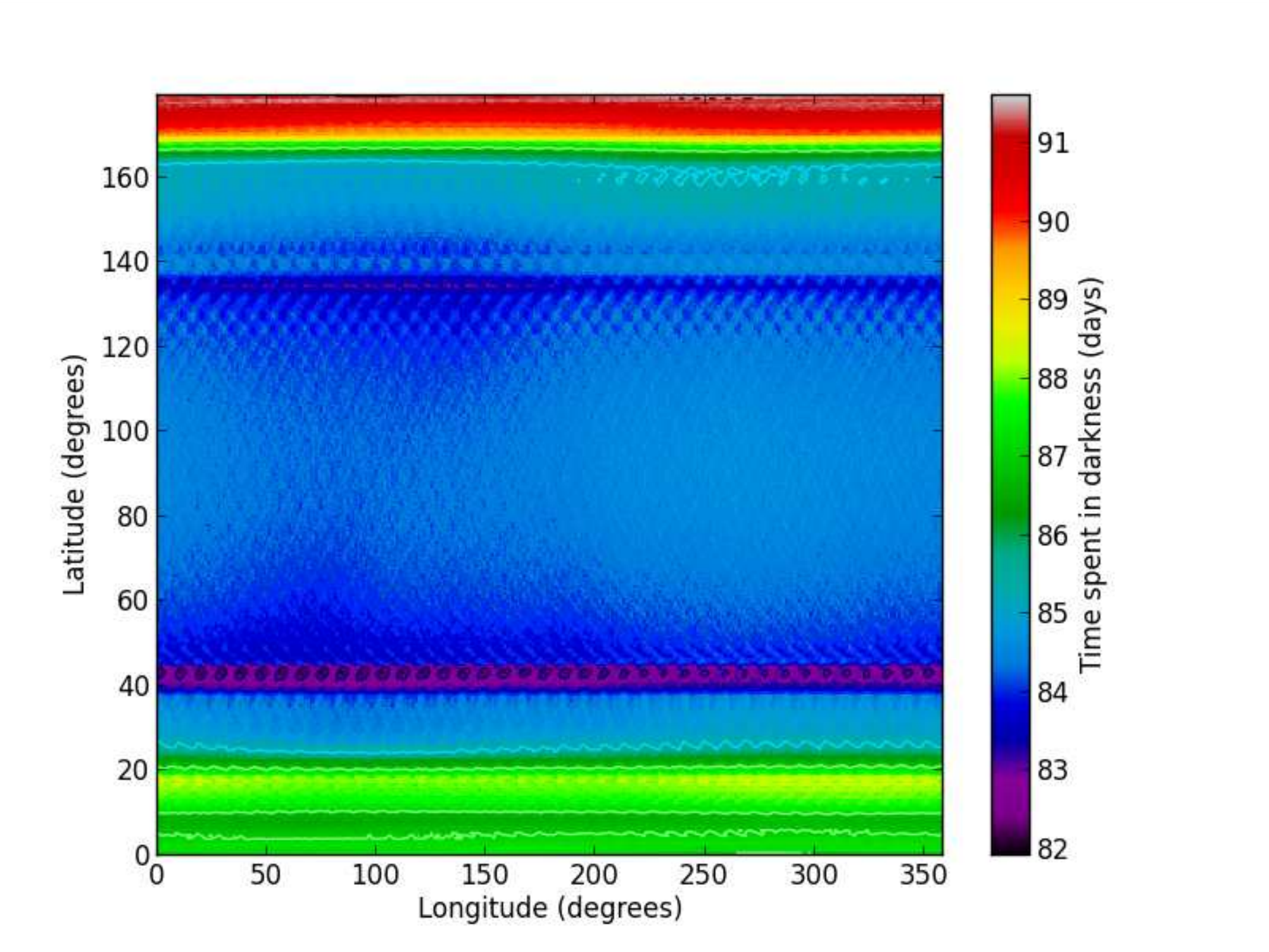} \\
\includegraphics[scale=0.4]{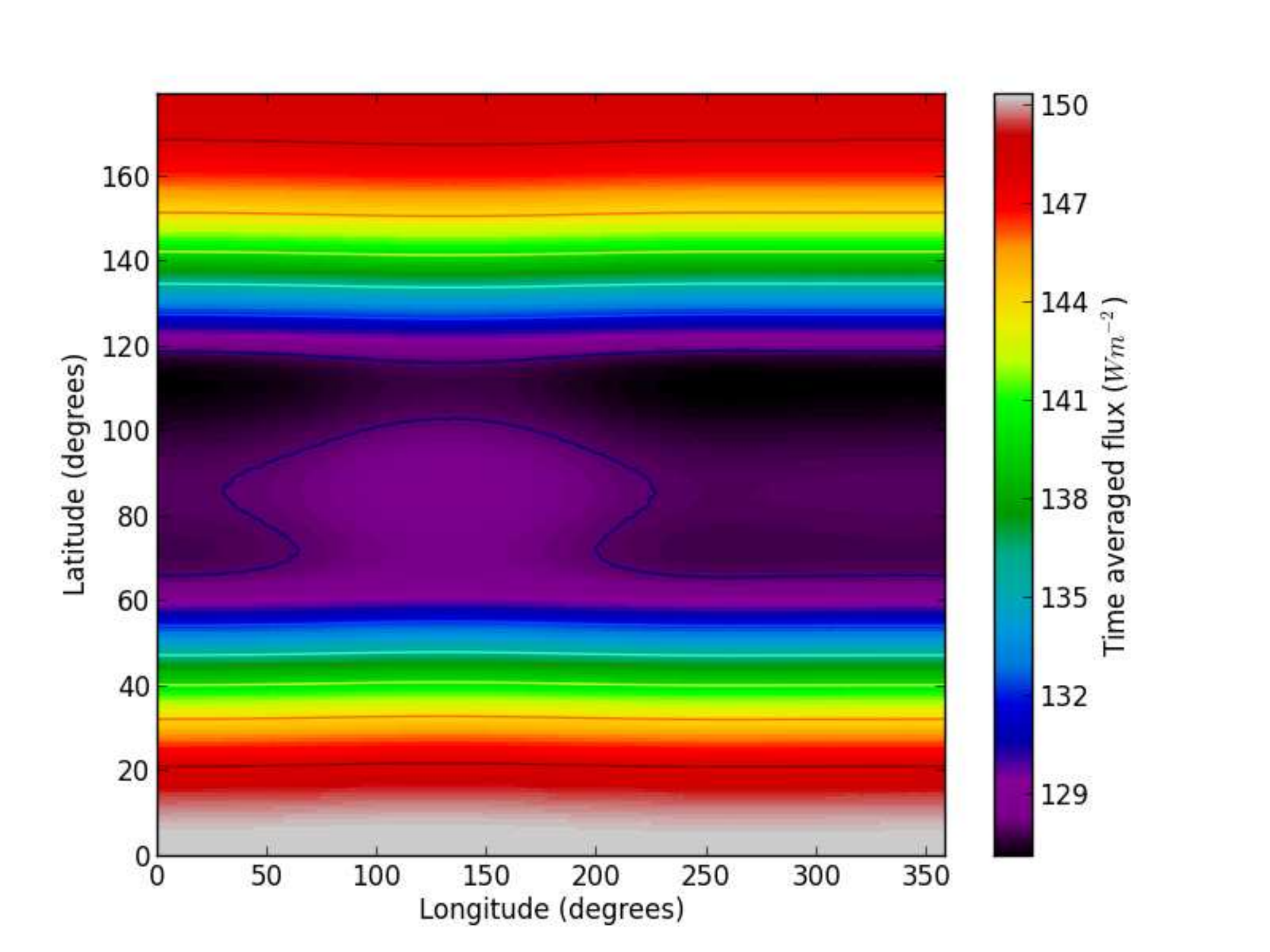} &
\includegraphics[scale=0.4]{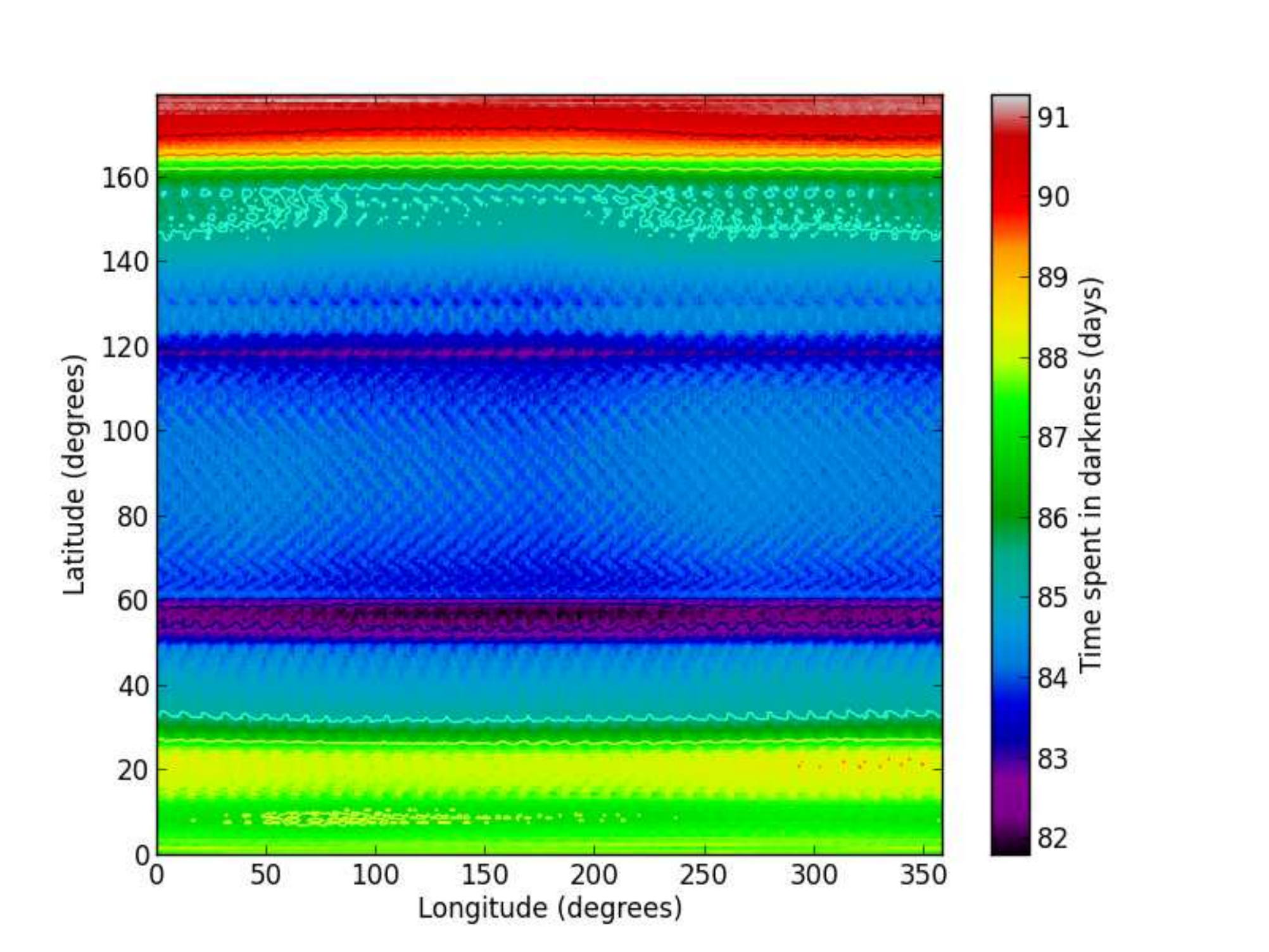} \\
\end{array}$
\caption{The effect of increasing orbital inclination.  Flux patterns (left column) and darkness patterns (right column)for a circumbinary planet in orbit around the Kepler-16 binary with obliquity $\delta_p=30^\circ$, eccentricity $e_p=0$, and (top row) orbital inclination $i_p=0$, (middle row) $i_p=15^\circ$, and (bottom row) $i_p=30^\circ$ \label{fig:inc_kep16}}
\end{figure*}

Kepler-16 presents a larger $a_{bin}$ and more equal mass ratio.  Figure \ref{fig:inc_kep16} shows there are some small differences in the presented flux and darkness features compared to Kepler-47.  The flux is distributed over a larger latitudinal range for $i_p=15^\circ$, and the eclipse timing effects which gave the south pole more flux for $\delta_p=60^\circ$, $i_p=0^\circ$ become apparent again for $\delta_p=30^\circ$, $i_p=30^\circ$.


\subsection{Sensitivity to the binary orbital phase}

\noindent Many of the flux patterns we have studied show strong longitudinal dependences, in particular where the substellar points are during the planet's periastron passage.  As $P_{orb,bin}$ and $P_{orb,p}$ are generally incommensurate, we should investigate how the relative orbital phases affect the resulting flux pattern.  The simplest probe of this effect is to change the longitude of periastron of the planet's orbit, set at 0$^\circ$ throughout.  As planet periastron occurs at $t=0$, the relative orbital phase is changed as a result.

\begin{figure*}
$\begin{array}{c}
\includegraphics[scale=0.4]{kepler16b_e05_ob60.pdf} \\
\includegraphics[scale=0.4]{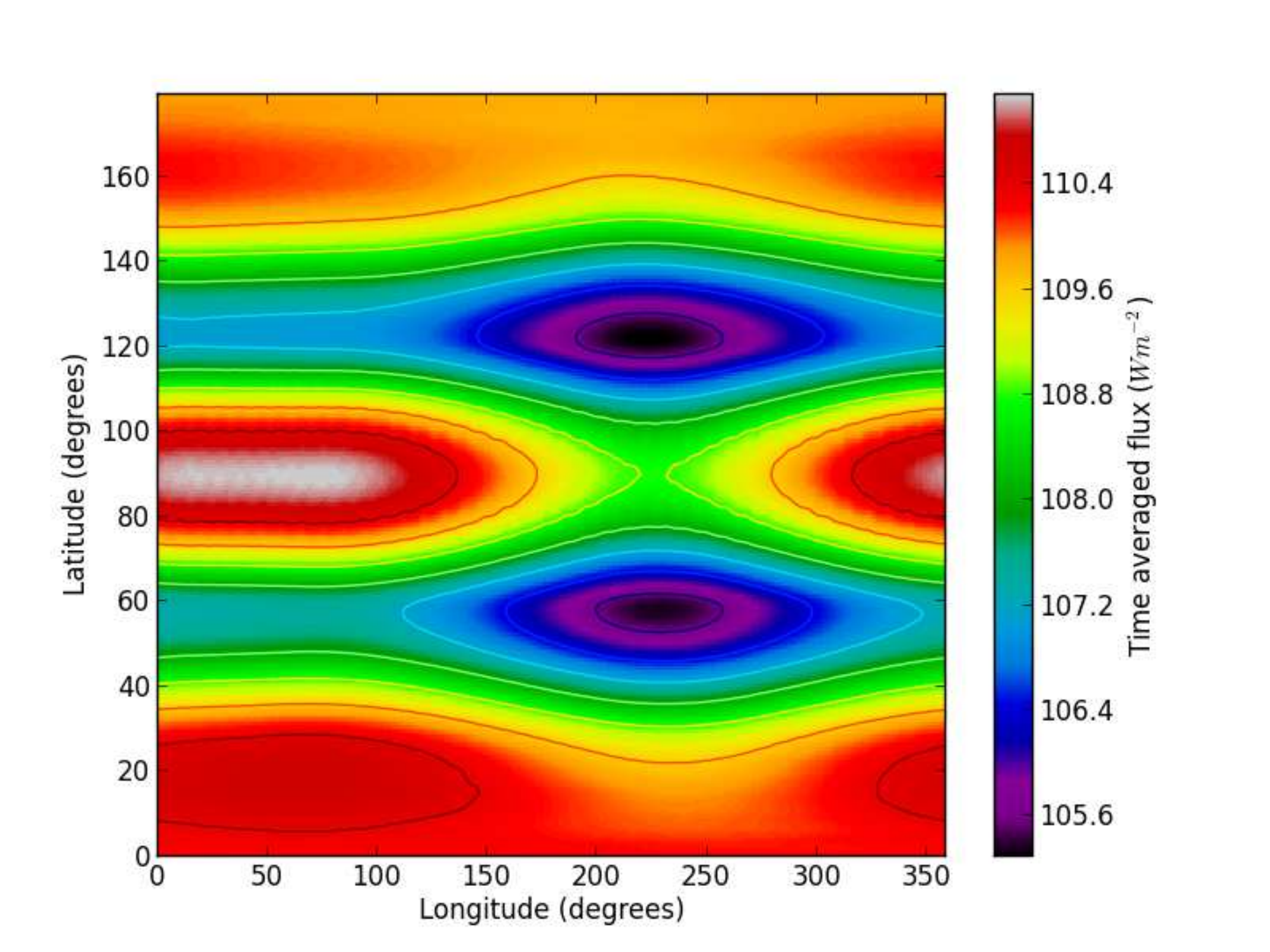} \\
\includegraphics[scale=0.4]{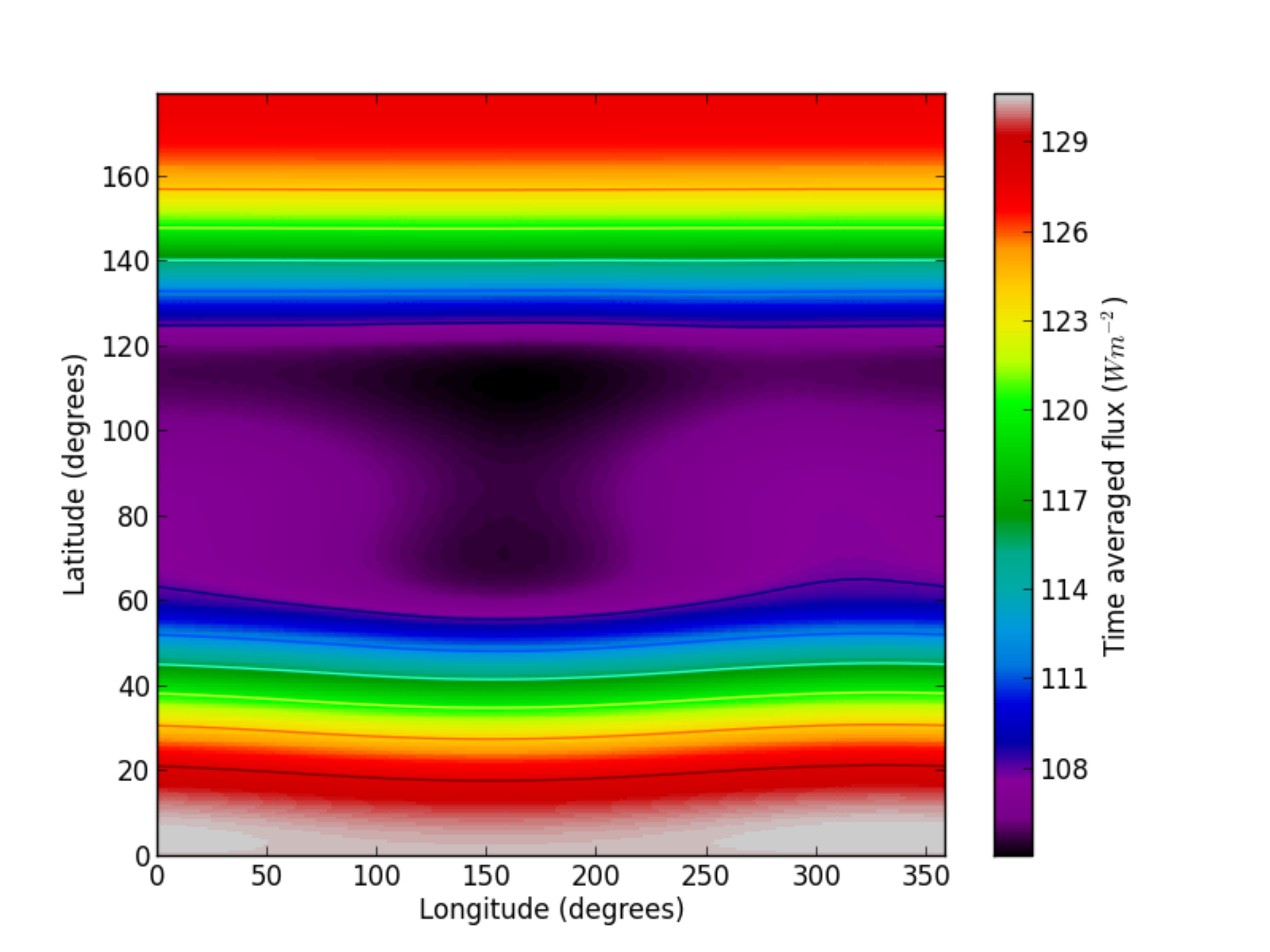} \\
\end{array}$
\caption{The effect of relative orbital phases.  Flux patterns for a circumbinary planet in orbit around the Kepler-16 binary with obliquity $\delta_p=60^\circ$, eccentricity $e_p=0.5$, and (top) longitude of periapsis of $0^\circ$, (middle row) longitude of periapsis of $5^\circ$, and (bottom) longitude of periapsis of $45^\circ$.  \label{fig:phase_kep16}}
\end{figure*}

Figure \ref{fig:phase_kep16} displays the flux patterns observed in the Kepler-16 system (for $e_p=0.5$, $\delta_p=60^\circ$), as the longitude of periapsis is increased.  As the relative orbital phase changes, the strong peak in flux at southern latitudes weakens, and finally disappears.  This sensitivity to phase will have important consequences for fluxes averaged over more than one orbit, as we will see in the following section.

\subsection{The persistence of patterns over many orbits}

\noindent The patterns seen thus far relate to one full orbit of the planet around the centre of mass.  As we have seen that the underlying cause of these patterns are in the relationship between the planet's orbital phase, the binary's orbital phase and their other respective orbital parameters, will such unusual patterns persist over many orbits, or will they ``average out'' into a morphology more consistent with a single star system?

\begin{figure*}
$\begin{array}{cc}
\includegraphics[scale=0.4]{kepler16b_dark_e02_ob30.pdf} &
\includegraphics[scale=0.4]{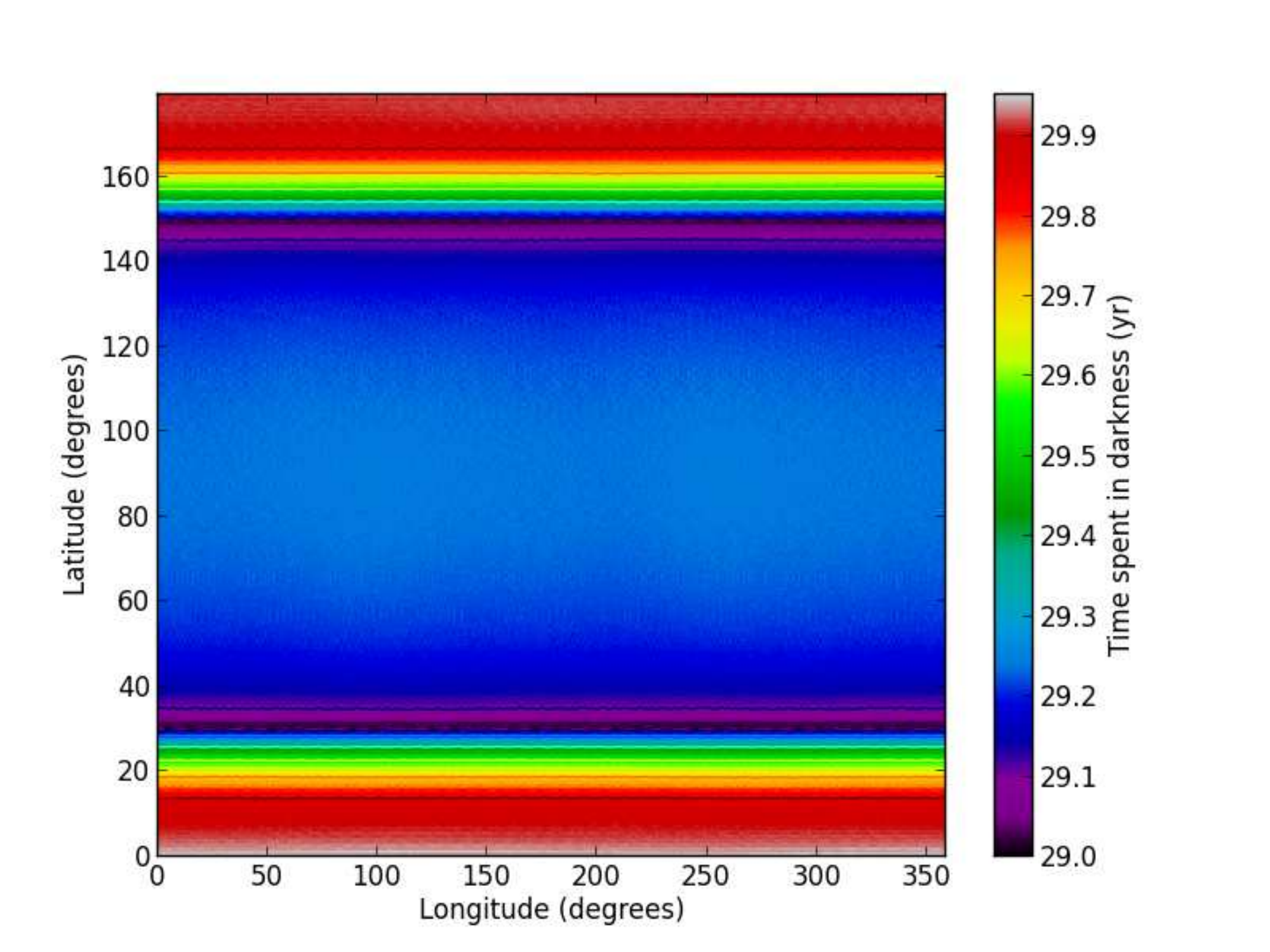}\\
\includegraphics[scale=0.4]{kepler47c_dark_e05_ob30.pdf} &
\includegraphics[scale=0.4]{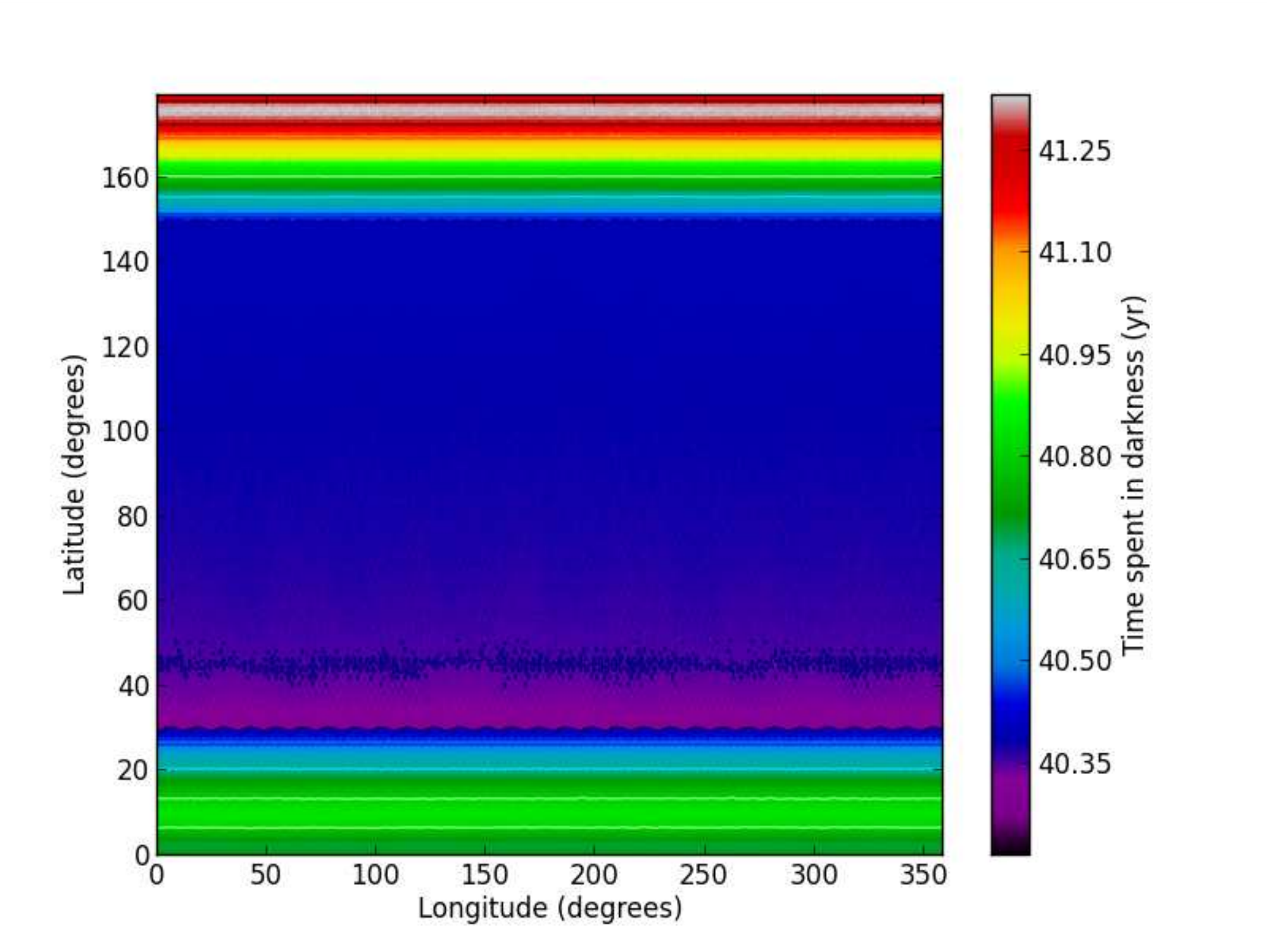} \\
\end{array}$
\caption{The persistence of patterns with time.  Top row: Darkness patterns for the Kepler-16b system, with $e_p=0.2$, $\delta_p=30^\circ$.  Bottom row: Darkness patterns for the Kepler-47 system, with $e_p=0.5$, $\delta_p=30^\circ$.  The left column displays results from one orbital period of the planet; the right column displays results from 100 orbital periods of the planet. \label{fig:manyorbits}}
\end{figure*}

Generally speaking, the flux patterns average out to a more uniform distribution, as the difference between the binary and planet's orbital longitudes precesses, allowing the stars' substellar points to move across all longitudes.  The darkness patterns, however, remain distinct.  Figure \ref{fig:manyorbits} compares the darkness patterns over 1 planetary orbital period to 100 planetary orbital periods for two cases.  The first is the Kepler-16b system, with $e_p=0.2$ and $\delta_p=30^\circ$; the second is the Kepler-47c system with $e_p=0.5$, $\delta_p=30^\circ$.  

\subsection{Orbital Stability, and the potential for spin-orbit resonances}

\noindent Until now, we have considered circumbinary planets with the same rotation period as the Earth, $P_{spin}=24$ h, but this is clearly not generally true, even for ``Earthlike'' planets around single stars.  As a consequence of total angular momentum conservation, tidal interactions between a host star and a planet can transfer spin angular momentum into orbital angular momentum.  Typically, the long term tidal evolution of exoplanets at close distances to their star results in circular orbits with synchronous rotation, where $P_{spin}=P_{orb,p}$, with obliquities generally locked into a Cassini state corresponding either to high or low values \citep{Peale1969,Henrard1987}.  If the initial orbit is eccentric, the planet can be captured into a $m:n$ spin-orbit resonance, where $mP_{spin} = nP_{orb,p}$, with $m,n$ integers.

Mercury is currently locked in a 3:2 spin orbit resonance with the Sun, thanks to its modest eccentricity of 0.206.  It can be shown that this capture was probable \citep{Correia2004,Dobrovolskis2007}, and stable over long timescales.  As such, spin-orbit resonances such as 3:2 may be a common feature of habitable planets around M stars.  As has been previously mentioned, the effects of resonances between $P_{spin}$ and $P_{orb,p}$ on flux patterns have been considered in detail for single star systems \citep{Dobrovolskis2007,Dobrovolskis2009,Dobrovolskis2013, Brown2014}.

To date, there has not been any in-depth dynamical study of spin-orbit resonances for circumbinary planets.  Some authors have considered the tidal interactions between the binary stars, and the resulting impact on hazardous flare activity \citep{Mason2013}, but none consider the spin-orbit evolution of the planet itself.  This may be a reasonable lapse, as the calculations for single stars suggest that spin-orbit evolution such as that seen for Mercury (and proposed for other exoplanets) only occurs if $a_p$ is sufficiently small.

In circumbinary planetary systems, there is a lower limit for $a_p$, underneath which planetary orbits are no longer dynamically stable \citep{Holman1999}

\begin{equation} 
a_p > a_{dyn} = a_{bin}\left(1.6 + 5.1 e_{bin} +4.12 \mu_{bin} -2.22 e^2_{bin} -4.27 \mu_{bin}e_{bin} - 5.09\mu^2_{bin} +4.61 \mu^2_{bin} e^2_{bin}\right),
\end{equation}

\noindent where $\mu_{bin} = \frac{M_2}{M_1+M_2}$.  The Kepler-47 system can be thought of as a solar-type system perturbed by the presence of the M type companion, so it may be reasonable to assume that tidal interaction calculations made for the Solar System, while inaccurate for this case, will give a sense of broad trends.  In the absence of a detailed calculation of the combined tidal interaction between both stars and any planet in the system, we can compare $a_{dyn}$ to the orbital semimajor axis of Mercury, which is 0.381 AU, and consider whether in the first instance Mercury's orbit would be stable, and leave the details of the interaction to later work.  In the case of Kepler-47, $a_{dyn} = 0.202 $ AU, so without the benefit of detailed calculation, we should consider the possibility that some planets could enter a spin-orbit resonance.  However, as in the case of the Solar System, it is likely that planets susceptible to spin-orbit resonances in the Kepler-47 system would also lie outside the habitable zone.

\begin{figure*}
$\begin{array}{cc}
\includegraphics[scale=0.25,angle=0,origin=c]{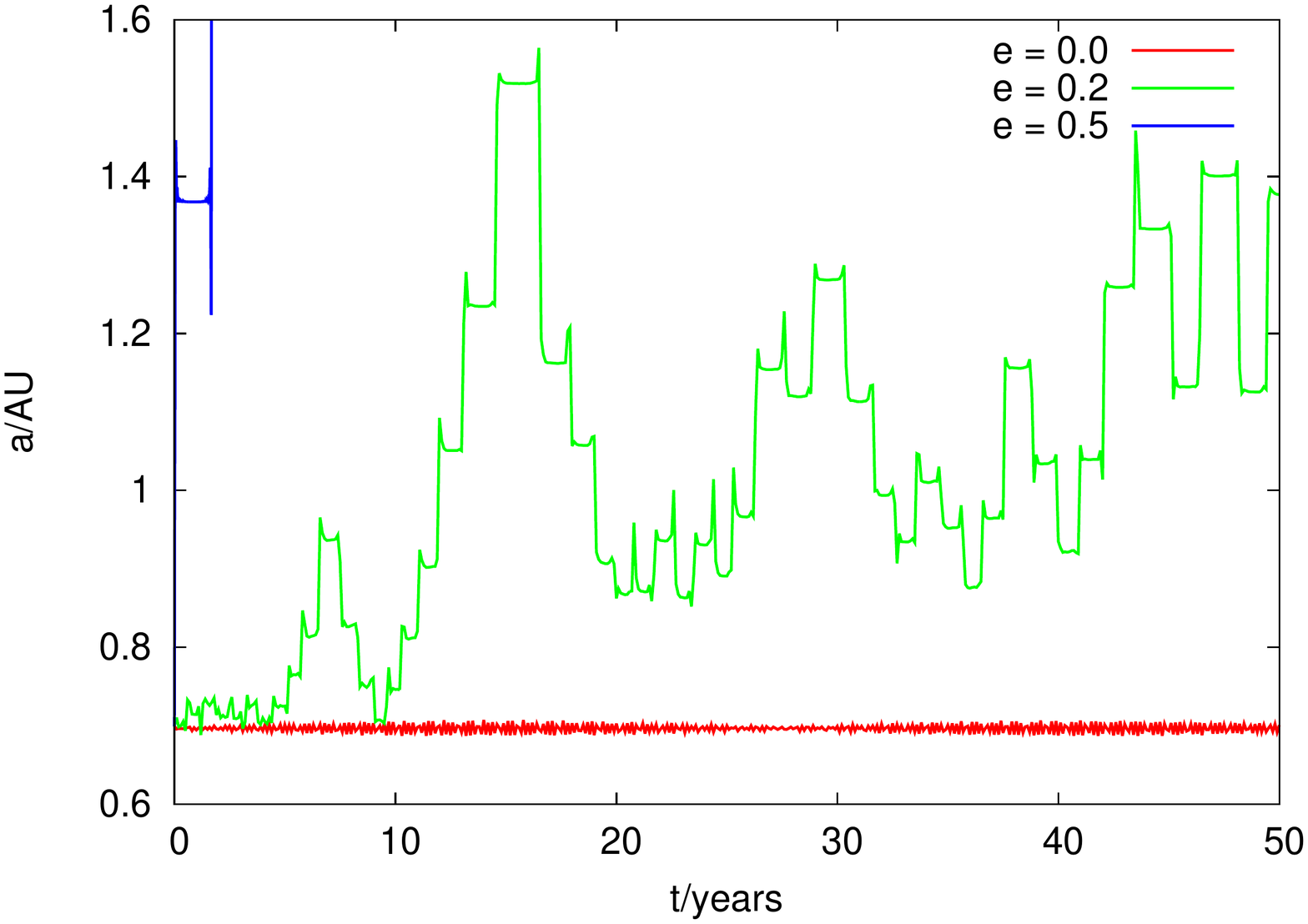} &
\includegraphics[scale=0.25,angle=0,origin=c]{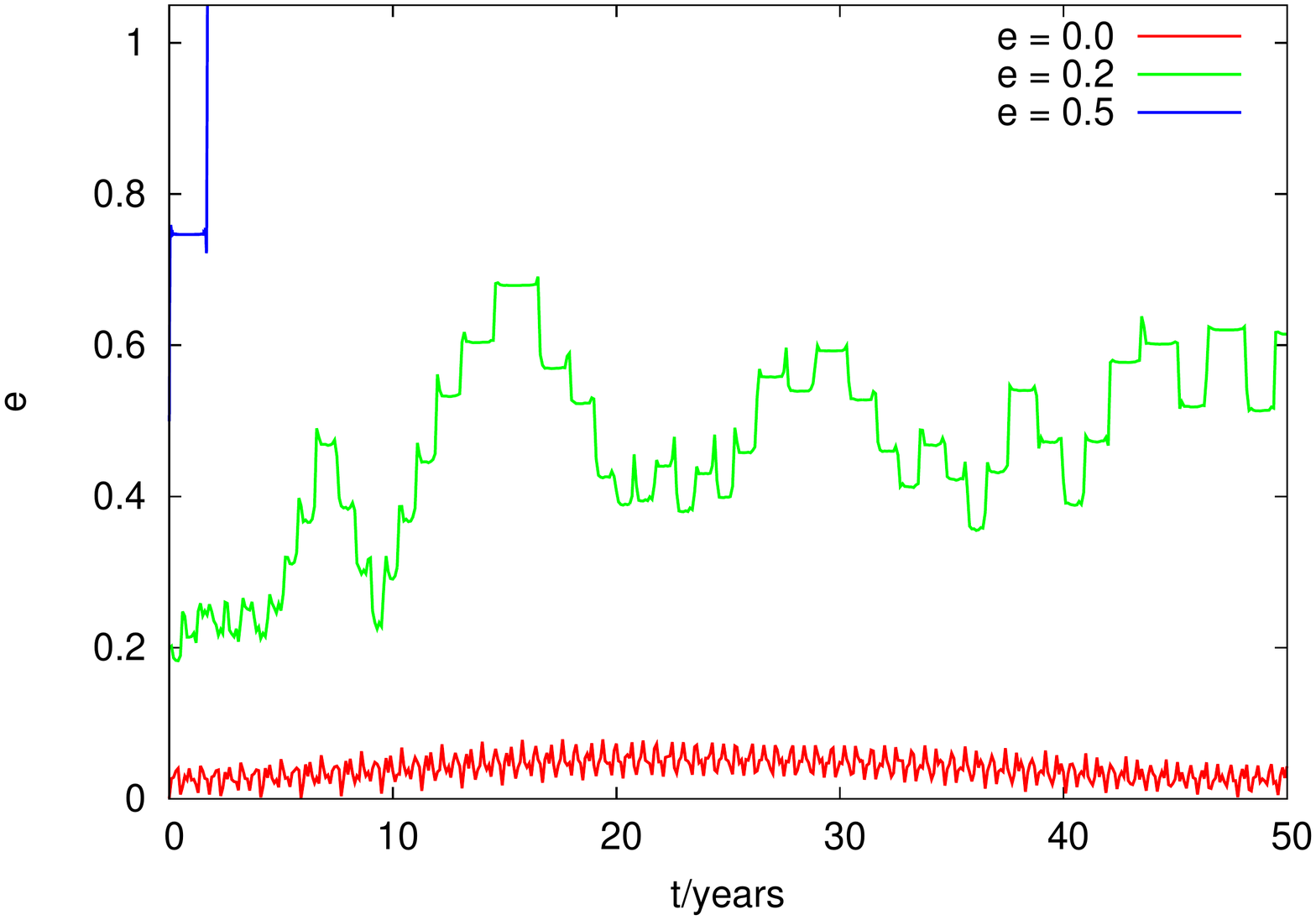} \\
\end{array}$
\caption{Orbital stability of planets orbiting the Kepler-16 binary under N-Body integration.  We show the semimajor axis (left), and eccentricity (right) for planets with initial eccentricities of 0 (red), 0.2 (green) and 0.5 (blue). The $e=0.5$ planet is ejected from the system after $\sim$ 2 years, whereas the $e=0.2$ planet is ejected after several hundred years.\label{fig:k16orbit}}
\end{figure*}

For Kepler-16, $a_{dyn} = 0.645$ AU, and most authors agree this means most of, if not all, of its habitable zone is orbitally unstable \citep{Liu2013}.  As we have already stated, eccentric orbits in the Kepler-16 system with Kepler-16b's semimajor axis are not stable, which we confirm via N-Body simulation (Figure \ref{fig:k16orbit}).  Given the low mass of both stars, it seems that planets cannot orbit sufficiently close to the binary to undergo significant tidal interactions, but once again we stress that this is at best an educated guess, and requires further work.

We should also note that circumbinary planets generally display rapid apsidal precession, behaviour that is confirmed both analytically and by N-Body simulation by other authors \citep{Doyle2011,Orosz2012, Leung2013}.  This precession will have important consequences for planetary seasons, and may also produce secular resonances, with corresponding effects on much longer timescales.

\subsection{Prospects for photosynthesis}

It seems clear that the patterns of both flux and darkness measured in this work could lead to patterns of rhythmicity that are unusual on the Earth (although we can appeal to some terrestrial analogues). The changing spectral quality reaching the planetary surface caused by the changing dominance of the radiation from the primary and secondary could potentially allow for a selection pressure that causes organisms to specialise in using particular regions of the spectrum, particularly as it shifts from short to longer wavelengths. 

\begin{figure*}
$\begin{array}{cc}
\includegraphics[scale=0.4]{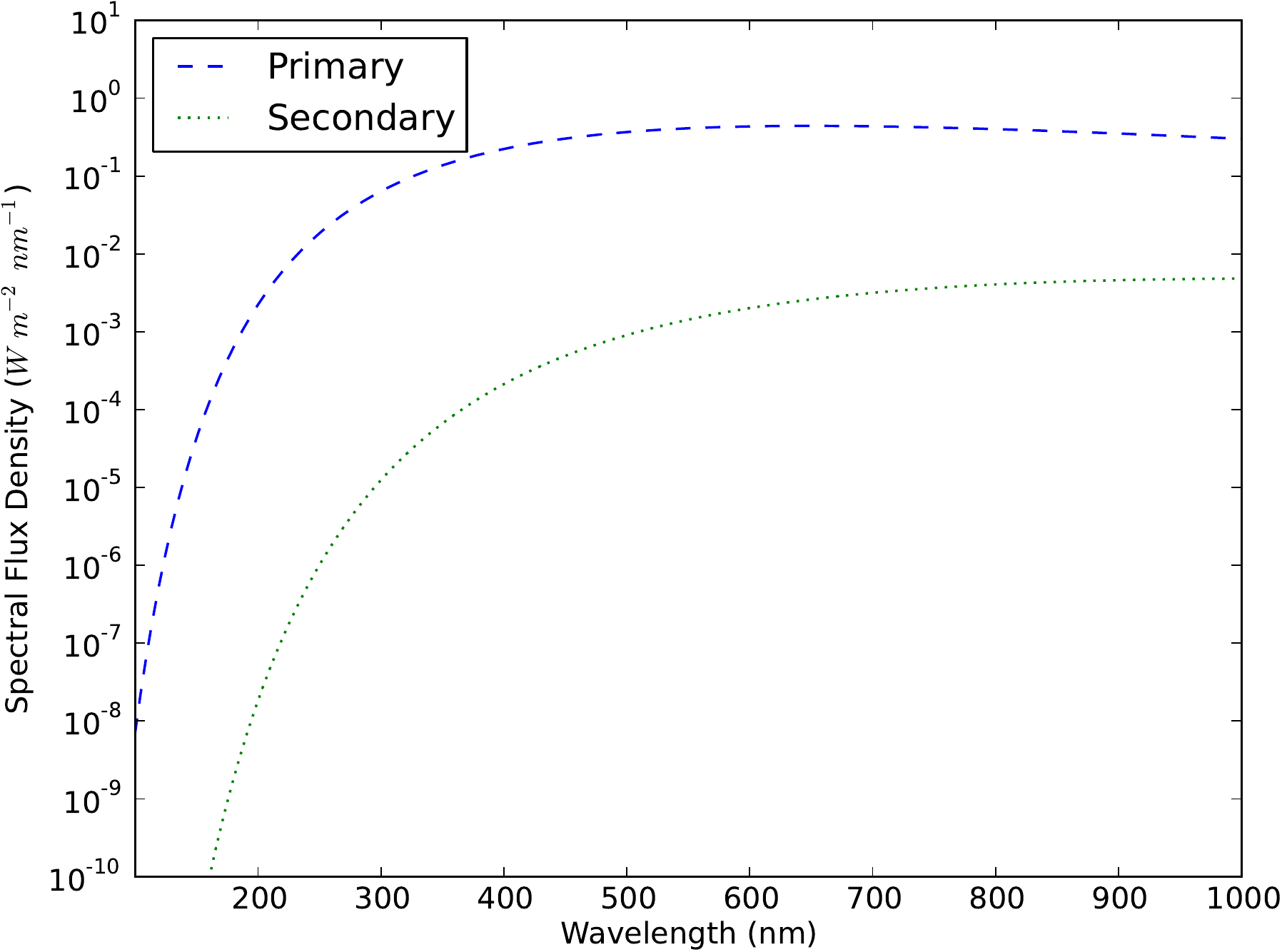} &
\includegraphics[scale=0.4]{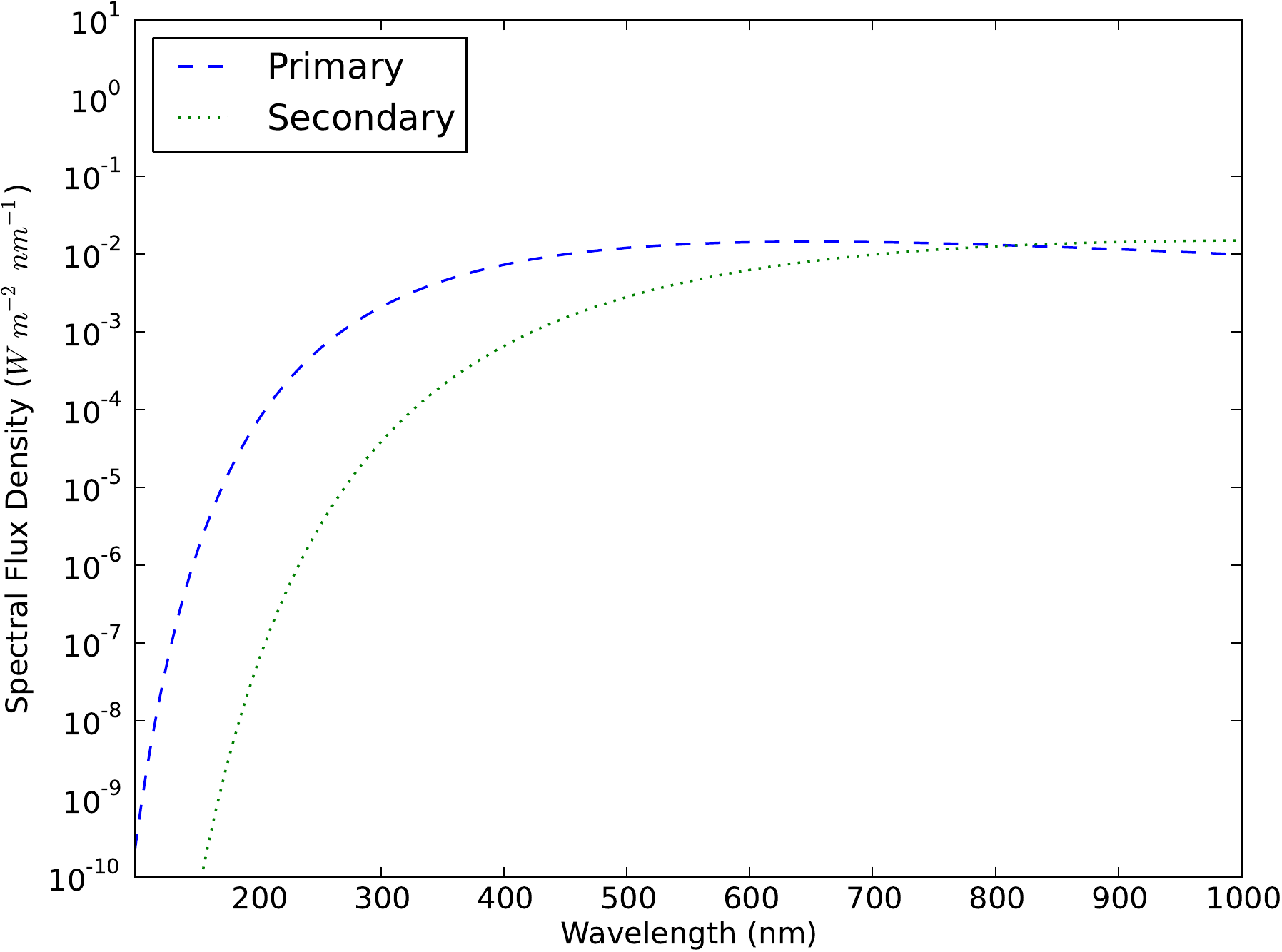} \\
\end{array}$
\caption{Spectral variation over the course of a (primary) day in the Kepler-16 system.  The left hand plot shows the spectral flux density when the primary is at maximum altitude (i.e. midday), and the right hand plot shows the same for when the primary is setting. \label{fig:spectrum_kep16}}
\end{figure*}

This spectral niche differentiation is seen on the Earth, across scales similar to that observed at different depths in the Earth’s oceans \citep{Stomp2007,Raven2007}. \citet{O'Malley2012a} considered photosynthesis on an Earth-like planet orbiting in the habitable zone of a close binary system comprising a G and an M star. That paper emphasises the possibility of a greater significance of photosynthetic organisms using radiation at wavelengths out to about 1000 nm than is the case on Earth, although the radiation environment of the Earth-like planet orbiting a G-M binary is dominated by the G star.  Figure \ref{fig:spectrum_kep16} shows the spectral flux density reaching the surface of a planet in the Kepler-16 system, orbiting at zero eccentricity and zero obliquity.  In the left plot, the primary is close to its maximum altitude in the sky (primary midday), and in the right plot it is close to the horizon (primary sunset).  Note that the secondary flux is increasing as the primary flux is decreasing, and exceeds the primary flux at infrared wavelengths during and after primary sunset.  The effects of such changes in infrared flux on photosynthesis depends on the extent to which global photosynthesis depends on photon absorption in the 700-1000 nm range; this is in turn a function of the global occurrence of infrared-absorbers and the depth of infrared-absorbing water overlying them \citep{Wolstencroft2002, Stomp2007, O'Malley2012a}.

The changing intensity of light and darkness on the planets discussed is analogous to the onset of lunar day, in which an additional light cycle with a lower frequency is superposed on the diurnal day/night cycle. On Earth, it has been found that animals can synchronise their activities to a great diversity of natural, geophysical rhythms, for example diurnal (24 hours), tidal (12.4 hours), semilunar (14.8 days), lunar (29.5 days), annual, seasonal, or photoperiodic (365 days), as well as longer cycles involving prime numbers of years (13 or 17) of reproduction related to predator satiation of the hatching organisms \citep{Sota2013}.  There is an evolutionary advantage of responding to stimuli which correlate with recurring environmental conditions. The lunar cycle is known to cause an enormous number of biological responses \citep{Endres2002}, for example the timing of breeding in marine organisms.  This is despite the moon being relatively faint (some 400,000 times fainter than the Sun in the optical).  It is true that some of these biological cycles are driven by tides, but we argue that biological rhythms driven by light cycles are likely to be present in organism behaviour of inhabited circumbinary planets, especially as the secondary light source is significantly brighter.  Given that the periodicity of light cycles in circumbinary systems can be both greater than and less than the planet's rotational period, we should expect that in biospheres of circumbinary planets, selection pressures will exist for biological cycles tuned to a variety of different rhythms, with the principal cycles linked to the period of the binary.

\section{Conclusions} \label{sec:conclusions}

We have calculated time-dependent surface maps of both flux and darkness on planets in circumbinary systems.  Using Kepler-16 and Kepler-47 as archetypes, we model the flux arriving from both stars as a function of planet latitude and longitude, given an initial Keplerian orbit with fixed elements, and a fixed spin obliquity. 

We identify patterns in both the flux and darkness that are unique to binary star systems.  These patterns exist as both changing spatial distributions and temporal fluctuations.  The spectral quality of radiation that might be considered photosynthetically useful also varies with time and surface position.  All the above variations have periods, amplitudes and phases that depend on the relative orbital phase between the stellar orbit and the planetary orbit, with timescales both much larger than the planetary orbital period and much smaller.

It is clear that if climate modelling of Earthlike circumbinary planets is to go beyond simple 1D approximations \citep{Forgan2014} towards more sophisticated 3D global circulation models (e.g. \citealt{Shields2014,Yang2014}) then these surface flux and darkness patterns will play a crucial role in determining both atmospheric physics and oceanic circulation (and the subsequent interactions thereof).

With such a wide variety of forcing timescales for photosynthesis, we conclude that any inhabited circumbinary planet will produce a biosphere rich in rhythms and cycles, determined by non-trivial relationships between the planet's orbit, the binary orbit, and the location of biomes on the planetary surface.  

\section*{Acknowledgments}

\noindent DF and CC gratefully acknowledge support from STFC grant ST/J001422/1. AM acknowledges the support of a STFC studentship. The University of Edinburgh is a charitable body, registered in Scotland, with registration number SC005336. The University of Dundee is a charitable body registered in Scotland, with registration number SC 15096.

\bibliographystyle{mn2e} 
\bibliography{Ptype_2Dflux}

\end{document}